\documentclass{article}

\usepackage{microtype}
\usepackage{graphicx}
\usepackage{subfigure}
\usepackage{booktabs} 

\usepackage[accepted]{icml2024}

\usepackage{amsmath}
\usepackage{amssymb}
\usepackage{mathtools}
\usepackage{amsthm}
\usepackage{lipsum}
\allowdisplaybreaks

\makeatletter
\def\addcontentsline#1#2#3{%
  \addtocontents{#1}{\protect\contentsline{#2}{#3}{\thepage}}}
\long\def\addtocontents#1#2{%
  \protected@write\@auxout
    {\let\label\@gobble \let\index\@gobble \let\glossary\@gobble}%
    {\string\@writefile{#1}{#2}}}
\makeatother

\usepackage{hyperref}
\definecolor{mydarkblue}{rgb}{0,0.08,0.45}
\hypersetup{ %
pdftitle={},
pdfauthor={},
pdfsubject={Proceedings of the International Conference on Machine Learning 2023},
pdfkeywords={},
pdfborder=0 0 0,
pdfpagemode=UseNone,
colorlinks=true,
linkcolor=mydarkblue,
citecolor=mydarkblue,
filecolor=mydarkblue,
urlcolor=mydarkblue,
pdfauthor={Anonymous Submission},
}

\newcommand{\ind}{\perp \!\!\!\!\perp} 
\usepackage{breqn}
\def\Xint#1{\mathchoice
{\XXint\displaystyle\textstyle{#1}}%
{\XXint\textstyle\scriptstyle{#1}}%
{\XXint\scriptstyle\scriptscriptstyle{#1}}%
{\XXint\scriptscriptstyle\scriptscriptstyle{#1}}%
\!\int}
\def\XXint#1#2#3{{\setbox0=\hbox{$#1{#2#3}{\int}$ }
\vcenter{\hbox{$#2#3$ }}\kern-.6\wd0}}
\def\dashint{\Xint-}

\usepackage[capitalize,noabbrev]{cleveref}
\theoremstyle{plain}
\newtheorem{theorem}{Theorem}[section]
\newtheorem{proposition}[theorem]{Proposition}
\newtheorem{lemma}[theorem]{Lemma}

\theoremstyle{definition}
\newtheorem{definition}[theorem]{Definition}
\newtheorem{assumption}[theorem]{Assumption}
\newtheorem{example}[theorem]{Example}
\theoremstyle{remark}
\newtheorem{remark}[theorem]{Remark}

\usepackage{bbm}
\usepackage{enumerate}
\usepackage{enumitem}
\usepackage[textsize=tiny]{todonotes}

\icmltitlerunning{Causal Discovery via Conditional Independence Testing with Proxy Variables}

\begin{document}

\twocolumn[
\icmltitle{Causal Discovery via Conditional Independence Testing with Proxy Variables}




\begin{icmlauthorlist}
\icmlauthor{Mingzhou Liu}{pku}
\icmlauthor{Xinwei Sun}{fdu}
\icmlauthor{Yu Qiao}{sjtu}
\icmlauthor{Yizhou Wang}{pku}
\end{icmlauthorlist}

\icmlaffiliation{pku}{C.S. Dep. Peking University, liumingzhou@stu.pku.edu.cn}
\icmlaffiliation{fdu}{Sch. of Data Sci., Fudan University}
\icmlaffiliation{sjtu}{C.S. Dep. Shanghai Jiao Tong Univeristy}

\icmlcorrespondingauthor{Xinwei Sum}{sunxinwei@fudan.edu.cn}

\icmlkeywords{causal discovery, conditional independence test, unobserved variables, proxy variables}

\vskip 0.3in
]



\printAffiliationsAndNotice{}  

\addtocontents{toc}{\protect\setcounter{tocdepth}{0}}

\begin{abstract}
    Distinguishing causal connections from correlations is important in many scenarios. However, the presence of unobserved variables, such as the latent confounder, can introduce bias in conditional independence testing commonly employed in constraint-based causal discovery for identifying causal relations. To address this issue, existing methods introduced proxy variables to adjust for the bias caused by unobserveness. However, these methods were either limited to categorical variables or relied on strong parametric assumptions for identification. In this paper, we propose a novel hypothesis-testing procedure that can effectively examine the existence of the causal relationship over continuous variables, without any parametric constraint. Our procedure is based on discretization, which under completeness conditions, is able to asymptotically establish a linear equation whose coefficient vector is identifiable under the causal null hypothesis. Based on this, we introduce our test statistic and demonstrate its asymptotic level and power. We validate the effectiveness of our procedure using both synthetic and real-world data. Code is publicly available at \url{https://github.com/lmz123321/proxy_causal_discovery}.
\end{abstract}

\section{Introduction}

Distinguishing causation from correlation is a fundamental task in various fields, such as economics, healthcare, and policy formulation. By avoiding spurious correlations caused by confounding bias or mediation bias, it facilitates drawing reliable conclusions and thereby making informed decisions. Under the causal sufficiency condition where all confounders are observed, one can directly conduct conditional independence testing \cite{zhang2012kernel} to examine the existence of causation, as employed by constraint-based methods for causal discovery \cite{glymour2019review}.

However, this procedure will suffer from nonignorable bias when \emph{unobserved variables} are present, such as latent confounders or mediators, as illustrated in Fig.~\ref{fig.proxy}. Typically, these variables refer to attributes that are difficult to measure (\emph{e.g.}, health status, living habits) but exhibit correlations to both variables within the specific pair of causal interests.

To address this challenge, recent studies in proximal causal learning introduced the proxy variable, which can be a noisy measurement \cite{kuroki2014measurement} or an observed descendant \cite{miao2018identifying} of the unobserved variable, to adjust for the bias caused by unobserveness. Examples include \cite{tchetgen2020introduction,deaner2021many,sverdrup2023proximal} that explored proxy variables for causal effect estimation; and \cite{ghassami2021proximal,dukes2023proximal} that studied proximal mediation analysis. However, most of these works focused on causal inference. The works that are mostly related to us are \cite{miao2018identifying} and \cite{miao2022identifying}, which introduced testing procedures to examine the null hypothesis of no causal relationship. However, their procedures were either limited to the discrete variables \cite{miao2018identifying} or relied on strong parametric assumptions to ensure the testing validity \cite{miao2022identifying}.

In this paper, we propose a novel hypothesis-testing procedure that can effectively identify the causal relationship between two continuous variables, without any parametric assumption. Our strategy is to find a proper discretization procedure such that, under the causal null hypothesis, the probability matrices over discretized variables satisfy a linear equation whose coefficient vector is identifiable. For this purpose, we first introduce a discretization procedure that, under the completeness assumption, makes the probability matrices over discrete variables satisfy rank regularities that are crucial for the identifiability of the coefficient vector. Then, we proceed to analyze the errors in the linear equation after discretization and show that they can be controlled to be arbitrarily small, given that the bin numbers for discretization are sufficiently large. Guided by the above analyses, we introduce our testing statistic based on the residue of linear regression and establish the asymptotic level and power of our test. We demonstrate the validity and effectiveness of our procedure on both synthetic and real-world data.

We summarize our contributions as follows:
\begin{enumerate}[noitemsep,topsep=1pt]
    \item We propose a proximal hypothesis testing based on discretization for identifying causal relations between continuous variables.
    \item We establish the asymptotic level and power of our procedure without any parametric assumptions. 
    \item We demonstrate the validity of our procedure on both synthetic and real-world datasets.
\end{enumerate}

\section{Related works}

\textbf{Proxy causal learning.} The proximal causal learning was first introduced by \cite{kuroki2014measurement} for causal inference with unobserved confounders, and was later studied in \cite{miao2018identifying, tchetgen2020introduction,deaner2021many,sverdrup2023proximal,ghassami2023partial} for causal effect identification and in \cite{mastouri2021proximal,ghassami2022minimax} for estimation. Recently, \cite{rojas2017causal, liu2023causal} leveraged the proxy variable for causal discovery; however, they respectively focused on causal orientation and time-series causal discovery, which are different from the problem considered in this paper. The works most relevant to ours are \cite{miao2018identifying,miao2022identifying}. In \cite{miao2018identifying}, the authors introduced a hypothesis testing procedure with the proxy variable to adjust for the bias introduced by latent confounders; however, their procedure is limited to discrete variables. For continuous variables, \cite{miao2022identifying} constructed a testing procedure by examining the existence of the solution of the integral equation. However, they rely on parametric assumptions to establish the asymptotic behavior of the test statistic under the null hypothesis. \textbf{In this paper}, we propose a novel testing procedure to effectively identify causal relationships between continuous variables, applicable to both parametric and non-parametric models.

\textbf{Conditional Independence (CI) testing.} CI testing is a fundamental step for constraint-based causal discovery. To effectively test CI, many works have been proposed. Examples include \cite{zhang2012kernel,doran2014permutation,pfister2018kernel,strobl2019approximate} that explored the kernel embedding and the Hilbert-Schimidt Independence Criterion (HSIC) for CI testing, \cite{runge2018conditional,mukherjee2020ccmi,ai2022testing} that used the mutual information as a measure to quantify the CI, and \cite{margaritis2005distribution,huang2010testing,neykov2021minimax,warren2021wasserstein} that introduced discretization-based procedures for CI testing. However, these works rely on the causal sufficiency condition that all conditional variables are observable. \textbf{As a contrast}, our testing procedure can effectively test CI in the presence of unobserved variables. 

\section{Preliminary}

In this section, we introduce the problem setup, notations, and background knowledge that is related to our work. 

\begin{figure}[t]
    \centering
    \includegraphics[width=.4\textwidth]{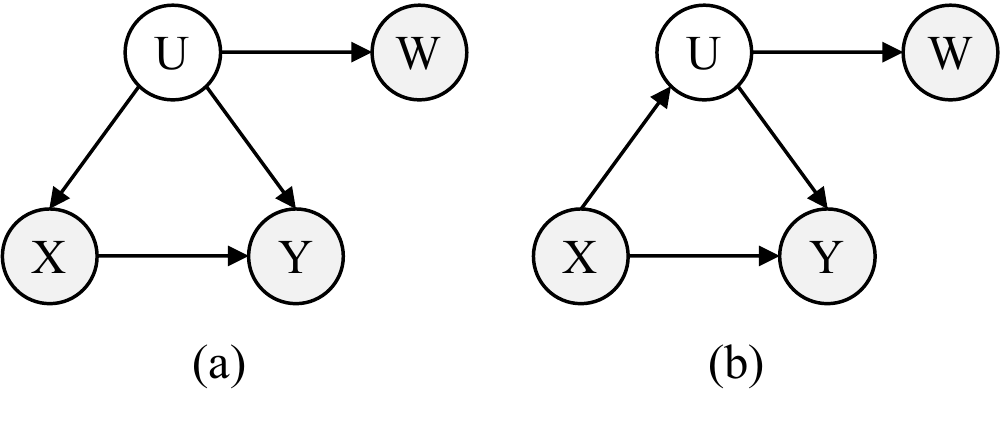}
    \caption{Causal diagrams illustrating causal discovery with proxy variables. (a) and (b) respectively represent the cases where $U$ is a latent confounder and a latent mediator. Note that our procedure is not restricted to these diagrams, but can apply to any scenario satisfying $X\ind W|U$ \cite{kuroki2014measurement}.}
    \label{fig.proxy}
    \vspace{-0.3cm}
\end{figure}

\textbf{Problem setup.} We consider the problem of identifying the causal relationship within a pair of variables $X$ and $Y$. The system also includes an unobserved variable $U$, which can be either a sufficient confounder or mediator between $X$ and $Y$. For this case, causal identification is generally not feasible, due to the confounding bias \cite{jager2008confounding} or the mediation bias \cite{hicks2011causal}. To adjust for the bias, we assume the availability of a proxy variable $W$ satisfying $X\ind W|U$\footnote{Our proposed procedure can be naturally applied to the scenario where multiple proxies are available, while the single-proxy scenario considered in this paper is more challenging.} \cite{kuroki2014measurement}. In practice, the proxy variable can be a noisy measurement \cite{kuroki2014measurement} or an observed descendant \cite{liu2023causal} of the unobserved variable $U$. We illustrate the relationships between $X$, $Y$, $U$, and $W$ in Fig.~\ref{fig.proxy}.

Under the Markovian and faithfulness assumptions \cite{pearl2009causality}, the problem of identifying the causal relationship between $X$ and $Y$ is equivalent to that of testing the causal null hypothesis $\mathbb{H}_0: X\ind Y|U$ with the proxy variable $W$. $\mathbb{H}_0$ means there is no causal relationship, and a rejection of $\mathbb{H}_0$ is, therefore, evidence in favor of causation.

\begin{remark}
    Our procedure can identify whether the causal relationship exists between $X$ and $Y$. For orientation, one can exploit prior knowledge such as temporal order, acyclicity constraint, or other properties such as asymmetry and stability \cite{tu2022optimal,duong2022bivariate}.
\end{remark}

\textbf{Notations.} Suppose $X,Y,U,W$ are random variables over the probability space $(\Omega,\mathcal{F},P)$, with domains $\mathcal{X},\mathcal{Y},\mathcal{U},\mathcal{W}$, respectively. We denote the probability law of $X$ as $\mathcal{L}_X:=P\circ X^{-1}$ and denote the cumulative density function (CDF) as $F_X$. For discrete variables, we denote their levels as $l_{X}, l_{Y}$, etc. We use the the column vector $P(X|y):=[p(x_1|y),...,p(x_{l_{X}}|y)]^\top$, the row vector $P(x|Y):=[p(x|y_1),...,p(x|y_{l_{Y}})]$, and the matrix $P(X|Y):=[P(X|y_{1}),...,P(X|y_{l_{Y}})]$ to represent their transition probabilities. For a continuous variable $X$, we denote $\{\mathcal{X}_i\}_{l_X}$ a disjoint partition of $\mathcal{X}$ with $l_X$ subspaces, such that $\cup_i^{l_X} \mathcal{X}_i=\mathcal{X}$ and $\mathcal{X}_i \cap \mathcal{X}_j =\emptyset$ for $i\neq j$. We thereby define the discretized version of $X$ as $\tilde{X}$, and denoting $\tilde{X}:= \tilde{x}_i$ to represent the event $X\in\mathcal{X}_i$ for any $i$. We denote $F(x|y)$ as the conditional CDF of $x$ given $Y=y$. Besides, we denote {\small $F(x|\boldsymbol{y}_{[i]}):=[F(x|y_1),...,F(x|y_i)]^\top$}, {\small $F(\boldsymbol{x}_{[i]}|y):=[F(x_1|y),...,F(x_i|y)]^\top$}, and {\small $F(\boldsymbol{x}_{[i]}|\boldsymbol{y}_{[i]}):=[F(x_1|\boldsymbol{y}_{[i]}),...,F(x_i|\boldsymbol{y}_{[i]})]$}, with $[i]:=\{1,...,i\}$. We let $A_1 \backslash A_2:=A_1\cap A_2^c$ for two sets $A_1, A_2$. We denote $n$ as the sample size. 


\textbf{The procedure described by \cite{miao2018identifying}.} For discrete variables, \cite{miao2018identifying} provided a hypothesis-testing procedure based on the proxy variable $W$\footnote{Their procedure was originally designed with two proxy variables $W$ and $Z$, satisfying $l_{X}l_{Z}>l_{W}$. Indeed, we can drop $Z$ and only use the single proxy $W$ if $l_{X}>l_{W}$.}. In details, they connected $\mathbb{H}_0:X\ind Y|U$ with the following equation:
\begin{equation}
    \underbrace{P(y|X)^\top}_{\text{response}} = \underbrace{P(W|X)^\top}_{\text{design matrix}} \{\underbrace{P(y|U) P(W|U)^{-1}}_{\text{coefficient}}\}^\top, 
    \label{eq.miao}
\end{equation}
by noticing that:
\begin{subequations}
\label{eq.cond-ind}
    \begin{align}
    P(y|X) &= P(y|U) P(U|X) \quad \mathbb{H}_0:X\ind Y|U, \\ 
    P(W|X) &= P(W|U) P(U|X) \quad X\ind W|U. 
\end{align}
\end{subequations}
Under the assumptions that the matrix $P(W|X)$ has full row rank and $P(W|U)$ is invertible, the linear system described by Eq.~\eqref{eq.miao} holds and the coefficient vector is identifiable under $\mathbb{H}_0$. In this regard, one can test $\mathbb{H}_0$ by investigating the residue via linear regression. 

While it has proven effective for discrete variables, extending such a hypothesis-testing procedure to continuous variables can be challenging. This is because, the conditional \textbf{independence} property encoded in Eq. (2), which is key to establishing the linear relation in Eq. (1), may not hold after discretization. Additionally, to ensure the \textbf{identifiability} of the coefficient vector, one should design a valid discretization procedure to guarantee the full row rank of ${P}(W | X)$ after discretization. In the next section, we propose a novel approach to address these challenges.

\vspace{-0.1cm}
\section{Methodology}
\vspace{-0.1cm}

In this section, we introduce a hypothesis-testing procedure based on discretization. At the core of our procedure lies finding a discretization procedure that establishs the identifiability of Eq.~(\ref{eq.miao}). 

To achieve this goal, in Sect.~\ref{sec.comp}, we first introduce a discretization procedure which, under the completeness assumption, can ensure the discretized $\tilde{X}$ and $\tilde{W}$ meet the rank regularity. Then, in Sect.~\ref{sec.dis-error}, we analyze the errors induced during the discretization. Under smoothness conditions regarding conditional distributions, we show that they can be effectively controlled to be arbitrarily small if the bin number is sufficiently large. Finally, based on the above analyses, we introduce our testing statistic and discuss the asymptotic level and power of the test in Sect.~\ref{sec.hypo-test}.

\subsection{Discretization under completeness}
\label{sec.comp}

In this section, we introduce a discretization procedure to establish the identifiability of Eq.~\eqref{eq.miao}. As mentioned previously, this involves finding discretization for $W$ and $X$ such that the resulted $P(\tilde{W}|\tilde{X})$ has full row rank. To begin, we first introduce the following completeness conditions that have been similarly adopted for continuous variables in causal inference \cite{miao2018identifying, miao2022identifying}, which ensures the existence of such a discretization.

\begin{assumption}[Completeness]
    We assume the conditional distributions $\mathcal{L}_{U|X}$ and $\mathcal{L}_{W|X}$ are complete. That is, for any bounded function $g$, we assume
    \begin{align*}
        \mathbb{E}[g(u)|x]=0 \,\, &\text{almost surely iff} \,\, g(u)=0 \,\, \text{almost surely}, \\
        \mathbb{E}[g(w)|x]=0 \,\, &\text{almost surely iff} \,\, g(w)=0 \,\, \text{almost surely}.
    \end{align*}
    \label{asm.comp-main}
\end{assumption}
\vspace{-.6cm}
To understand why we need Asm.~\ref{asm.comp-main} in our method, note that when $W$ is discrete, the completeness of $\mathcal{L}_{W|X}$ means the conditional probability matrix $P(W|X)$ has full row rank, as required in the procedure for discrete variables \cite{miao2018identifying}. When $W$ is continuous, this condition enables us to find a discretization such that $P(\tilde{W}|\tilde{X})$ has full row rank. Further, when combined with the completeness of $\mathcal{L}_{U|X}$, we can demonstrate the existence of discretization for $U$ such that the matrix $P(\tilde{W}|\tilde{U})$ is invertible, which is also required for the identifiability of Eq.~\eqref{eq.miao}.

The following example shows that Asm.~\ref{asm.comp-main} can hold for a wide range of Addictive Noise Models (ANMs).

\begin{example}[ANM with completeness]
    Suppose that $X,U,W$ satisfy the ANM, \emph{i.e.}, $X=g_X(U)+N_X, W=g_W(U)+N_W$, with $N_X,N_W$ being the noises. If $g_X$ is invertible with a non-zero derivative, and $N_X,N_W$ have non-zero characterization functions, then the conditional distributions $\mathcal{L}_{U|X}$ and $\mathcal{L}_{W|X}$ are both complete.
    \label{exam.anm.completeness}
\end{example}

\begin{remark}
    We also provide an example where $U$ is a mediator. Please refer to Appx.~\ref{app-sec.example-comp} for details.
\end{remark}

The completeness conditions have been similarly assumed in causal hypothesis testing and causal inference, with the difference that our condition is not restricted to square-integrable functions. For example, the completeness of $\mathcal{L}_{U|X}$ was assumed by \cite{miao2018identifying,miao2022identifying}\footnote{See Condition~(ii) in \cite{miao2018identifying} (by dropping $Z$) and Asm.~3 in \cite{miao2022identifying}.}. Besides, a discrete version of the completeness of $\mathcal{L}_{W|X}$, \emph{i.e.}, $P(W|X)$ has full row rank, was required by the procedure described by \cite{miao2018identifying}. Indeed, in \cite{miao2018identifying}, they also assumed the completeness of $\mathcal{L}_{X|W}$\footnote{See Condition~(iii), where $z$ should be replaced with $x$ to induce variation for identification.}. However, this assumption is required primarily for causal effect estimation, which therefore is not required in this paper.

Next, we show that, under Asm.~\ref{asm.comp-main}, we can find a discretization procedure such that the $P(\tilde{W}|\tilde{X})$ has full row rank and the matrix $P(\tilde{W}|\tilde{U})$ is invertible.

\begin{proposition}
    Suppose Asm.~\ref{asm.comp-main} holds. Then, for any discretization $\tilde{W}$ of $W$, there exists a discretization $\tilde{X}$ of $X$ such that the matrix $P(\tilde{W}|\tilde{X})$ has full row rank. Similarly, there also exists a discretization $\tilde{U}$ of $U$ such that the matrix $P(\tilde{W}|\tilde{U})$ is invertible.
    \label{prop.rank-dis-main}
\end{proposition}

\begin{remark}
    We leave the proof to Appx.~\ref{app-sec.discrete}. Roughly speaking, we first show that given $\tilde{W}$ after discretizing $W$, the functions $F(x|\tilde{w}_1),...,F(x|\tilde{w}_{l_{{W}}})$ are linear independent under the completeness condition of $\mathcal{L}_{W|X}$. Then we prove with induction that there exists $\tilde{x}_1,...,\tilde{x}_{l_{{X}}}$ with $l_{{X}} > l_{{W}}$, such that $\left[F(\boldsymbol{\tilde{x}}_{[l_{{X}}]}|\tilde{w}_1),...,F(\boldsymbol{\tilde{x}}_{[l_{{X}}]}|\tilde{w}_{l_{{W}}})\right]$ are linear independent, implying the full column rank of $P(\tilde{X}|\tilde{W})$ and therefore the full row rank of $P(\tilde{W}|\tilde{X})$. 
\end{remark}

\begin{remark}
    While we cannot discretize $U$ due to its unobserved nature, this is not necessary as we do not need to directly compute $P(\tilde{W}|\tilde{U})$ in the linear regression. Instead, we only require it to be invertible with a proper discretization of $U$, as ensured by Prop.~\ref{prop.rank-dis-main}, to achieve identifiability of the coefficient vector in Eq.~\eqref{eq.miao}.
\end{remark}

Inspired by Prop.~\ref{prop.rank-dis-main}, we introduce our discretization procedure in Alg.~\ref{alg.dis}. The algorithm mainly contains three steps. First, in line 1, we uniformly discretize $W$ into $l_{{W}}$ bins to obtain $\tilde{W}$. Indeed, the choice of discretization for $W$ is arbitrary, as indicated by Prop.~\ref{prop.rank-dis-main}. Next, in lines 2-8, we discretize $X$ to construct $\tilde{X}^\prime$ that also contains $l_{{W}}$ bins, such that $P(\tilde{X}^\prime|\tilde{W})$ and therefore $P(\tilde{W}|\tilde{X}^\prime)$ is invertible. Finally, we add another $(l_{{X}}-l_{{W}})$ bins to $\tilde{X}^\prime$ to obtain $\tilde{X}$ in line 9. In this regard, $P(\tilde{W}|\tilde{X})$ has full row rank.

In the second step (lines 2-8), we iteratively search for $x_i$ such that the matrix $M_i:=F(\boldsymbol{x}_{[i]}|\boldsymbol{\tilde{w}}_{[i]})$ is invertible given $x_1,...,x_{i-1}$. This can be achieved by noting that the functions $F(x|\tilde{w}_1),...,F(x|\tilde{w}_{l_{{W}}})$ are linear independent under the completeness condition. Once we have the invertibility of $M_i$, we have $P(\tilde{X}|\tilde{W}):=\{P(\tilde{x}_i|\tilde{w}_j)\}_{i,j}$ is invertible, with $P(\tilde{x}_i|\tilde{w}_j):= F(x_i|W \in \mathcal{W}_j)-F(x_{i-1}|W \in \mathcal{W}_j)$ and $\mathcal{X}_i:=[x_{i-1},x_i)$ $\left( (-\infty,x_1) \text{ for $i=1$}\right)$. To find such a $x_i$, note that $\mathrm{det}(M_i)$ can be expanded as:
\vspace{-.1cm}
\begin{align*}
    \mathrm{det}(M_i)(x_i) = \sum_{j=1}^i (-1)^{1+j} d_j F(x_i|\tilde{w}_j),
\end{align*}
where $d_j:=\mathrm{det}(F(\boldsymbol{x}_{[i-1]}|\tilde{\boldsymbol{w}}_{[i]\backslash j}))$. By induction, there exists a non-zero $(i-1)$ dimensional minor among $\{d_j\}$. Since $F(x|\boldsymbol{\tilde{w}}_{[i]})$ are linear independent, there exists $x_i$ s.t. $\mathrm{det}(M_i)(x_i) \neq 0$ and therefore $F(\boldsymbol{x}_{[i]}|\tilde{\boldsymbol{w}}_{[i]})$ is invertible.

\begin{algorithm}[t]
   \caption{Discretizing $W$ and $X$.}
   \label{alg.dis}
\begin{flushleft}
\textbf{Input:} Continuous data $\{x_i,w_i\}_{i=1}^n$; $l_{{X}}> l_{{W}}$. \\
\textbf{Output:} {\small $\{\mathcal{X}_i\}_{i=1}^{l_{{X}}}, \{\mathcal{W}_j\}_{j=1}^{l_{{W}}}$} s.t. {\small $P(\tilde{W}|\tilde{X})$} has full row rank.
\end{flushleft}
\begin{algorithmic}[1]
    \STATE Discretize $\mathcal{W}$ into $l_{{W}}$ uniform bins to obtain $\{\mathcal{W}_j\}_{j=1}^{l_{{W}}}$.
    \FOR{$i=1:l_{{W}}$}
        \FOR{$j=1:i$}
            \STATE Estimate $d_j=\mathrm{det}(F(\boldsymbol{x}_{[i-1]}|\tilde{\boldsymbol{w}}_{[i]\backslash j}))$.
        \ENDFOR
        \STATE Search for $x_i$ s.t. $\sum_{j=1}^i (-1)^{1+j} d_j F(x_i|\tilde{w}_j) \neq 0$.
        \STATE Set $\mathcal{X}_i = [x_{i-1},x_i)$.
    \ENDFOR
    \STATE Discretize $\mathcal{X}\backslash [x_1,x_{l_{{W}}})$ into $(l_{{X}}-l_{{W}})$ uniform bins to obtain $\{\mathcal{X}_i\}_{i=l_{{W}}+1}^{l_{{X}}}$.
\end{algorithmic}
\end{algorithm}

\textbf{Choosing the bin number.}
    The bin number is a trade-off between the discretization error and the estimation of discrete probabilities. While a larger bin number effectively controls discretization error (as explained in the next section), it may lead to inaccurate probability estimation due to insufficient samples in each bin. To balance this trade-off, we follow the heuristics proposed by \cite{dougherty1995supervised} to set the bin number to $\left \lfloor \log(n) \right \rfloor$.

\subsection{Discretization error analysis}
\label{sec.dis-error}

In this section, we discuss the discretization error control. Given $\mathbb{H}_0:X\ind Y|U$ and $X\ind W|U$, let $\tilde{y}$ represent the event $Y\in\tilde{\mathcal{Y}}$ for some $\tilde{\mathcal{Y}}\subset \mathcal{Y}$, our goal is to have:
\vspace{-.1cm}
\begin{subequations}
\label{eq.dis-ind}
\begin{align}
    P(\tilde{y}|\tilde{X})&=P(\tilde{y}|\tilde{U}) P(\tilde{U}|\tilde{X}) + E_y, \\
    P(\tilde{W}|\tilde{X})&=P(\tilde{W}|\tilde{U}) P(\tilde{U}|\tilde{X}) + E_w,
\end{align}
\end{subequations}
with the error terms
\begin{align*}
    E_y(i) &=  \sum\nolimits_{j}\{P(\tilde{y}|\tilde{x}_i,\mathcal{U}_j)-P(\tilde{y}|\mathcal{U}_j)\}P(\mathcal{U}_j|\tilde{x}_i), \\
    E_w(i) &=  \sum\nolimits_{j}\{P(\tilde{W}|\tilde{x}_i,\mathcal{U}_j)-P(\tilde{W}|\mathcal{U}_j)\}P(\mathcal{U}_j|\tilde{x}_i),
\end{align*}
being controlled to zeros asymptotically for all $i$. In this regard, we can establish Eq.~(\ref{eq.miao}) asymptotically. 


We first consider the case where $\mathcal{U}$ is a compact space, then extend the results to $\mathcal{U}=\mathbb{R}^d$  ($d \in \mathbb{Z}_+$). For a compact $\mathcal{U}$, we can partition $\mathcal{U}$ into sufficiently many bins $\{\mathcal{U}_j\}_{l_U}$, such that in each small bin $\mathcal{U}_j$, the errors $|P(\tilde{y}|\tilde{x}_i,U\in \mathcal{U}_j)-P(\tilde{y}|U\in \mathcal{U}_j)|$ and $|P(\tilde{W}|\tilde{x}_i,U\in \mathcal{U}_j)-P(\tilde{W}|U\in \mathcal{U}_j)|$ can be effectively controlled. The key to this property is the following smoothness condition, which ensures the conditional distributions do not vary too much in each small bin $\mathcal{U}_j$.

\begin{assumption}[TV smoothness]
\label{asm.tv-smooth-main}
    We assume that the maps $u\mapsto \mathcal{L}_{Y|U=u}$ and $u\mapsto \mathcal{L}_{W|U=u}$ are Lipschitz continuous wrt the Total Variation (TV) distance. That is, we assume there exist $L_Y$ and $L_W$ such that for any $u,u^\prime \in \mathcal{U}$:
    \begin{align*}
        \mathrm{TV}(\mathcal{L}_{Y|U=u},&\mathcal{L}_{Y|U=u^\prime}) \leq L_Y |u-u^\prime|, \\
        \mathrm{TV}(\mathcal{L}_{W|U=u},&\mathcal{L}_{W|U=u^\prime}) \leq L_W |u-u^\prime|,
    \end{align*}
    with $\mathrm{TV}(\mathcal{L}_1,\mathcal{L}_2):=\sup_{F\in\mathcal{F}} |\mathcal{L}_1(F)-\mathcal{L}_2(F)|$.
\end{assumption}

To understand the role of Asm.~\ref{asm.tv-smooth-main} in controlling the error $|P(\tilde{y}|\tilde{x}_i,U\in \mathcal{U}_j)-P(\tilde{y}|U\in \mathcal{U}_j)|$ , note that for the small bin $\mathcal{U}_j$ and any $u\in\mathcal{U}_j$, Asm.~\ref{asm.tv-smooth-main} means that the distance
\begin{equation*}
    \mathrm{TV}(\mathcal{L}_{Y|U\in \mathcal{U}_j},\mathcal{L}_{Y|U=u})
\end{equation*}
is small. Similarly, it also means that the distance
\begin{equation*}
    \mathrm{TV}(\mathcal{L}_{Y|X\in\mathcal{X}_i,U\in \mathcal{U}_j},\mathcal{L}_{Y|X\in\mathcal{X}_i,U=u})
\end{equation*}
is small. Since we have $\mathcal{L}_{Y|X\in\mathcal{X}_i,U=u}=\mathcal{L}_{Y|U=u}$ under $\mathbb{H}_0: X\ind Y|U$, by applying the triangle inequality, we can control the distance $\mathrm{TV}(\mathcal{L}_{Y|X\in\mathcal{X}_i,U\in \mathcal{U}_j},\mathcal{L}_{Y|U\in \mathcal{U}_j})$ and hence control the term $|P(\tilde{y}|\tilde{x}_i,U\in \mathcal{U}_j)-P(\tilde{y}|U\in \mathcal{U}_j)|$.

The following example shows that Asm.~\ref{asm.tv-smooth-main} can hold for a wide range of ANMs.

\begin{example}[ANM with TV smoothness]
    Suppose that $Y,U,W$ satisfy the ANM, \emph{i.e.}, $Y=g_Y(U)+N_Y, W=g_W(U)+N_W$, with $N_Y, N_W$ being the noises. If $g_Y, g_W$ are differentiable, and the probability densities of $N_Y, N_W$ have absolutely integrable derivatives, then the maps $u\mapsto \mathcal{L}_{Y|U=u}$ and $u\mapsto \mathcal{L}_{W|U=u}$ are Lipschitz continuous with respect to TV distance.
    \label{exam.tv-smooth}
\end{example}

Under Asm.~\ref{asm.tv-smooth-main}, we can control the discretization error as long as the length of each bin is sufficiently small.

\begin{proposition}
\label{prop.control-err-main}
    Suppose that $\mathcal{U}$ is a compact space, and that $\{\mathcal{U}_j\}_{l_{U}}$ is a partition of $\mathcal{U}$ satisfying
    \vspace{-.1cm}
    \begin{equation}
        len(\mathcal{U}_j)\leq \frac{1}{2}min\{\epsilon/L_Y,\epsilon/L_W\}, \label{eq.small-bin}
    \end{equation}
    for every bin $\mathcal{U}_j$. Then, under Asm.~\ref{asm.tv-smooth-main} and $\mathbb{H}_0$, for any $i$, we have that:
    \vspace{-.1cm}
    \begin{align*}
        &|P(\tilde{y}|\tilde{x}_i,U\in\mathcal{U}_j) - P(\tilde{y}|U\in\mathcal{U}_j)| \leq \epsilon, \\
        &|P(\tilde{W}|\tilde{x}_i,U\in\mathcal{U}_j) - P(\tilde{W}|U\in\mathcal{U}_j)| \leq \epsilon,
    \end{align*}
    and therefore $E_y(i)\leq \epsilon, E_w(i) \leq \epsilon$ for any $i$.
\end{proposition}

According to Prop.~\ref{prop.control-err-main}, when $\mathcal{U}$ is a compact space, we can (implicitly) partition $\mathcal{U}$ into sufficiently refined bins, by setting the bin number large enough. However, when $\mathcal{U}$ is unbounded, it is intractable to partition it into finite bins with finite length. Therefore, to extend the above result to $\mathcal{U}=\mathbb{R}^d$, we additionally assume $\mathcal{L}_{U|X}$ to be tight, meaning that for large values of $u$'s, $p(u|x)$ is close to zero. The formal definition of a tight distribution is given below: 

\begin{definition}[Tight distribution] The marginal distribution $\mathcal{L}_U$ is said to be tight if $\forall \epsilon>0, \exists t_0, \forall t\geq t_0,\mathcal{L}(|U|\geq t)\leq \epsilon$. The conditional distribution $\mathcal{L}_{U|X}$ is said to be uniformly tight in $[a,b]$ if $\forall \epsilon>0, \exists t_0, \forall t\geq t_0, \forall x\in [a,b],\mathcal{L}(|U|\geq t|x)\leq \epsilon$.
\label{def.tight-main}
\end{definition}

Now we introduce the tightness condition below:
\begin{assumption}
    We assume the conditional distribution $\mathcal{L}_{U|X}$ is uniformly tight for any compact interval $[a,b]$.
    \label{asm.tight-main}
\end{assumption}

To understand how Asm.~\ref{asm.tight-main} can be applied in the case when $\mathcal{U}=\mathbb{R}^d$, note that we can write $p(\tilde{y}|\tilde{x}_i)$ as: 
\begin{align*}
    p(\tilde{y}|\tilde{x}_i) & = \underbrace{p\left(\tilde{y} | |U|\leq t_0,\tilde{x}_i\right)p\left(|U| \leq t_0  |\tilde{x}_i\right)}_{\text{compact}} \\
    & \quad \quad \quad \quad + \underbrace{p\left(\tilde{y} | |U|>t_0,x\right)p\left(|U|>t_0 |\tilde{x}_i\right)}_{\text{non-compact}}.
\end{align*}
As the $t_0$ goes to infinity, the ``non-compact" term approaches 0 and hence the error terms $\{E_y(i)\}_i, \{E_w(i)\}_i$ are only accounted for by the ``compact" term. 

The following example shows that Asm.~\ref{asm.tight-main} can easily hold when the distribution of the noise is tight.
\begin{example}[ANM with tightness]
    Suppose that $X$ and $U$ satisfy the ANM, \emph{i.e.}, $X=g_X(U)+N_X$, with $N_X$ being the noise. If $g_X$ is invertible and the distribution of $N_X$ is tight\footnote{E.g., distributions from the exponential family.}, then $\mathcal{L}_{U|X}$ is uniformly tight in any compact interval $[a,b]$ with $-\infty < a < b < \infty$.
    \label{exam.anm.tight}
\end{example}

Under Asm.~\ref{asm.tight-main}, we have the following result for controlling the error induced by the non-compact part of $\mathcal{U}$.

\begin{lemma}
    Suppose $\mathcal{L}_{U|X}$ is uniformly tight in $[a,b]$, and the map $x\mapsto\mathcal{L}_{U|X=x}$ is $L_U$-Lipschitz continuous wrt TV distance. Suppose $\{\mathcal{X}_i\}_{l_X}$ is a partition of $[a,b]$ satisfying
    \begin{equation}
        len(\mathcal{X}_i)\leq \epsilon/(2L_U). \label{eq.small-bin-2}
    \end{equation}
    Then, $\forall\epsilon>0, \exists t_0$ such that $P(\Vert U \Vert_2 > t_0 | \tilde{x}_i)<\epsilon$ for any $i$.
    \label{lem.tight-dis-err}
\end{lemma}

Combining Prop.~\ref{prop.control-err-main}, we then have the following result for discretization error control when $\mathcal{U}=\mathbb{R}^d$:

\begin{proposition}
    Suppose $\mathcal{U}=\mathbb{R}^{d}$ and assumptions in Prop.~\ref{prop.control-err-main} and Lem.~\ref{lem.tight-dis-err} hold. Denote $\{\mathcal{U}_j\}_{l_{U}}$ and $\{\mathcal{X}_i\}_{l_X}$ as partitions of $[-t_0,t_0]$ and $[a,b]$, respectively, s.t. (\ref{eq.small-bin}) and (\ref{eq.small-bin-2}) hold. Then, for the corresponded discretized variables $\tilde{U},\tilde{X}$, it holds that $E_y(i)\leq \epsilon$ and $E_w(i) \leq \epsilon$ for any $i$.
    \label{prop.dis-err unbd}
\end{proposition}

\vspace{-0.1cm}
\subsection{Hypothesis test}
\label{sec.hypo-test}
\vspace{-0.1cm}

In this section, we present our hypothesis-testing procedure. Specifically, we adopt the test statistic proposed by \cite{miao2018identifying} and analyze its asymptotic distribution under the null hypothesis. Additionally, we investigate the behavior of the statistic under the alternative hypothesis and establish the asymptotic power of our test.

We first introduce our test statistic. The idea is that we can assess whether $\mathbb{H}_0$ is correct by looking at the residue of the least-square solution of Eq.~\eqref{eq.miao}. If $\mathbb{H}_0$ is correct, then the linear relation in Eq.~\eqref{eq.miao} holds, hence the least-square residue should be close to zero. Specifically, we let
\begin{equation*}
    q:=P(\tilde{y}|\tilde{X})^{\top}, \,\,\, Q:=P(\tilde{W}|\tilde{X}).
\end{equation*}
Using the Maximum Likelihood Estimation (MLE)\footnote{Please refer to Appx.~\ref{app-sec.mle} for details.}, we can obtain estimators $(\hat{q},\hat{Q})$ satisfying that:
\begin{align*}
    n^{1/2}(\hat{q} - q) &\to N(0,\Sigma) \,\, \text{in distribution}, \\
    \hat{Q}\to Q, \hat{\Sigma} &\to \Sigma \,\, \text{in probability,} 
\end{align*}
with $\hat{\Sigma}, \Sigma$ both positive definite. Then, we have the least-square residue of regressing $\hat{\Sigma}^{-1/2} \hat{q}$ on $\hat{\Sigma}^{-1/2} \hat{Q}$ to be:
\begin{equation*}
    \xi := \{I - \hat{\Sigma}^{-1/2} \hat{Q}^\top (\hat{Q} \hat{\Sigma}^{-1} \hat{Q}^\top)^{-1} \hat{Q} \hat{\Sigma}^{-1/2}\} \hat{\Sigma}^{-1/2} \hat{q}.
\end{equation*}
Following \cite{miao2018identifying}, we adopt the squared residue $T:=n\xi^\top \xi$ as our test statistic. 

Next, we discuss the asymptotic distribution of $T$. Suppose that $W$ and $X$ are respectively discretized into $l_{{W}}= l$ and $l_{{X}}=l+r$ bins, with $r$ a positive number. Then, as $n,l$ both go to infinity, the statistic $T_{n,l}$ follows an asymptotic chi-squared distribution $\chi^2_r$, where the degree of freedom $r$ is determined by the additional levels that $\tilde{X}$ has beyond $\tilde{W}$. Formally, denote $z_{1-\alpha}$ as the $(1-\alpha)$-th quantile of $\chi^2_r$, we have the following result:

\begin{theorem}
    Suppose that Asm.~\ref{asm.comp-main},~\ref{asm.tv-smooth-main},~\ref{asm.tight-main} hold, then under $\mathbb{H}_0$, it holds that:
    \begin{equation*}
        T_{n,l} \to \chi_{r}^2 \,\, \text{in distribution}
    \end{equation*}
    as $n,l\to \infty$. As a result, for any $\alpha\in (0,1)$, we have:
    \begin{equation*}
        \lim_{n,l\to\infty} P(T_{n,l}> z_{1-\alpha}) = \alpha.
    \end{equation*}
    That means, our procedure has uniform asymptotic level $\alpha$.
\label{thm.control.type-i}
\end{theorem}

In general, we cannot expect that the test has power against alternatives to $\mathbb{H}_0$ without further restrictions on the alternative distribution. This is analogous to other conditional independence tests \cite{zhang2012kernel,runge2018conditional,christgau2023nonparametric} that are used for causal discovery. Therefore, to establish the power of our test, we put the following restriction on the alternative distribution:

\begin{assumption}
    We consider the local alternative $\mathbb{H}_1$, such that the following integral equation has no solution:
    \begin{equation}
        p(y|x)=\int h(y,w)p(w|x)dw.
    \label{eq.int-eq}
    \end{equation}
    \label{asm.int-eq-main}
\end{assumption}
\vspace{-.2cm}
\begin{remark}
    Empirically, we find that Asm.~\ref{asm.int-eq-main} can hold in many cases under $X\not\ind Y|U$, as shown in Appx.~\ref{app-sec.int-eq}. Besides, we can prove that $\mathbb{H}_1$ is indeed a local alternative, that is, there exists distributions satisfying $X\not\ind Y|U$ but do not satisfy Asm.~\ref{asm.int-eq-main}. See Appx.~\ref{app-sec.int-eq} for details.
    \label{rem.type-ii-cond}
\end{remark}

To understand the role of Asm.~\ref{asm.int-eq-main} in controlling the type II error, we first note that the residues in $T$ should deviate from zero under the alternative model, which requires that Eq.~\eqref{eq.miao} has no solution. Intuitively, this means the variability of $P(\tilde{y}|\tilde{X})$ can not be fully explained $P(\tilde{W}|\tilde{X})$, which then corresponds to Asm.~\ref{asm.int-eq-main} for continuous variables. On the other hand, we can show in Appx.~\ref{app-sec.proof-type-ii} that the solution in Eq.~\eqref{eq.int-eq} exists under $\mathbb{H}_0$. Therefore, we can employ it to effectively distinguish $\mathbb{H}_0$ from the alternative.

Under this assumption, we can establish the asymptotic power for our testing procedure.  
\begin{theorem}
    Suppose that Asm.~\ref{asm.comp-main},~\ref{asm.tv-smooth-main},~\ref{asm.tight-main},~\ref{asm.int-eq-main} hold, then for any $\alpha\in (0,1)$, it holds that:
    \vspace{-.2cm}
    \begin{equation*}
        \lim_{n, l\to\infty}P(T_{n,l}>z_{1-\alpha})=1.
    \end{equation*}
    In other words, the limiting power of our test is one.
    \vspace{-.3cm}
    \label{thm.type-ii-main}
\end{theorem}
\begin{figure*}[htbp]
     \centering
    \includegraphics[width=0.95\textwidth]{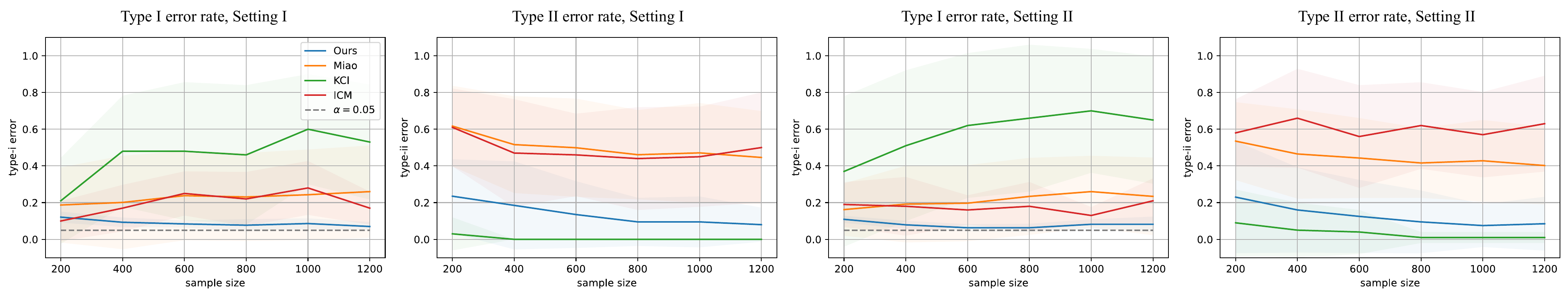}
    \caption{Type I and type II error rates of our testing procedure and baseline methods. Note that for a valid testing procedure, the type I error should be close to the significant level $\alpha$ (the dashed line), and the type II error should be close to zero.}
    \label{fig.cmp-baselines}
\end{figure*}

\section{Experiment}

In this section, we apply our method to synthetic data and causal discovery on a real-world dataset.

\textbf{Compared baselines.} We compare our method to the following baselines: \textbf{i)} the Kernel-based Conditional Independence test (KCI) \cite{zhang2012kernel} that tested the null hypothesis of $X\ind Y|W$ using kernel matrices; \textbf{ii)} the Integrated Conditional Moment test (ICM) that tested whether the conditional moment equation $\mathbb{E}[Y-h(W)|X]$ has a solution; and \textbf{iii)} the procedure described by \cite{miao2018identifying} (Miao for simplicity) that was designed for causal discovery over discrete variables using the proxy variable.

\textbf{Evaluation metrics.} We report the rate of type I and type II errors. The type I error is the mistaken rejection of a null hypothesis when it is true, and the type II error means the failure to reject a null hypothesis when it is false. Ideally, for a valid testing procedure, the type I error rate should be close to the significance level $\alpha$, and the type II error rate should be close to zero.

\textbf{Implementation details.} We set the significance level $\alpha$ to 0.05. We set the bin numbers of our method to $l_{{X}}=14,$ $l_{{W}}=12$, respectively. For KCI and ICM, we adopt the implementations provided in the $\mathrm{causallearn}$\footnote{\url{https://github.com/py-why/causal-learn}} and $\mathrm{mliv}$\footnote{\url{https://github.com/causal-machine-learning-lab/mliv}} packages, respectively. For the procedure described by Miao, we implement the R code released in the paper and set $l_{{X}}=3, l_{{W}}=2$ by default.

\subsection{Synthetic data}

\textbf{Data generation.} We consider two settings for data generation, using the confounding graph in Fig.~\ref{fig.proxy} (a) and the mediation graph in Fig.~\ref{fig.proxy} (b), respectively. For both settings, we follow the ANM to generate $V$, \emph{i.e.}, $V = f_V(\mathbf{PA}_V) + N_V$, for each $V \in \{X,Y,U,W\}$, with $\mathbf{PA}_V$ and $N_V$ denoting the associated parent nodes and the noise, respectively. For each $V$, the function $f_V$ is randomly chosen from $\{\mathrm{linear},\mathrm{tanh},\mathrm{sin},\mathrm{sigmoid}\}$, and the noise distribution is random selected from $\{\mathrm{Gaussian},\mathrm{uniform},\mathrm{exponential},\mathrm{gamma}\}$. To remove the effect of randomness, we repeat the process $20$ times. For each time, we generate $100$ replications under $\mathbb{H}_0$ and $\mathbb{H}_1$, respectively, to record the type I and type II error rates.

\textbf{Type I and type II error rates.} In Fig.~\ref{fig.cmp-baselines}, we report the performance of our testing procedure and baseline methods. As shown, the type I error rate of our method approximates the desired level $\alpha=0.05$, while other methods exhibit considerably higher rates. Besides, our type II error rate is getting close to 0 as $n$ increases. To explain, note that conditioning on the proxy variable $W$ of KCI cannot eliminate the confounding/mediation bias. The higher type I error rate of ICM may come from its limitation in hypothesis testing for only parametric models, as well as the estimation error. Besides, this procedure suffers from a large type II error rate. Finally, the additional errors of Miao's procedure may come from the violation of the linear equation in Eq.~\eqref{eq.miao} under $\mathbb{H}_0$ arising from coarse discretization.

\begin{figure}[tbp]
    \centering
    \includegraphics[width=.46\textwidth]{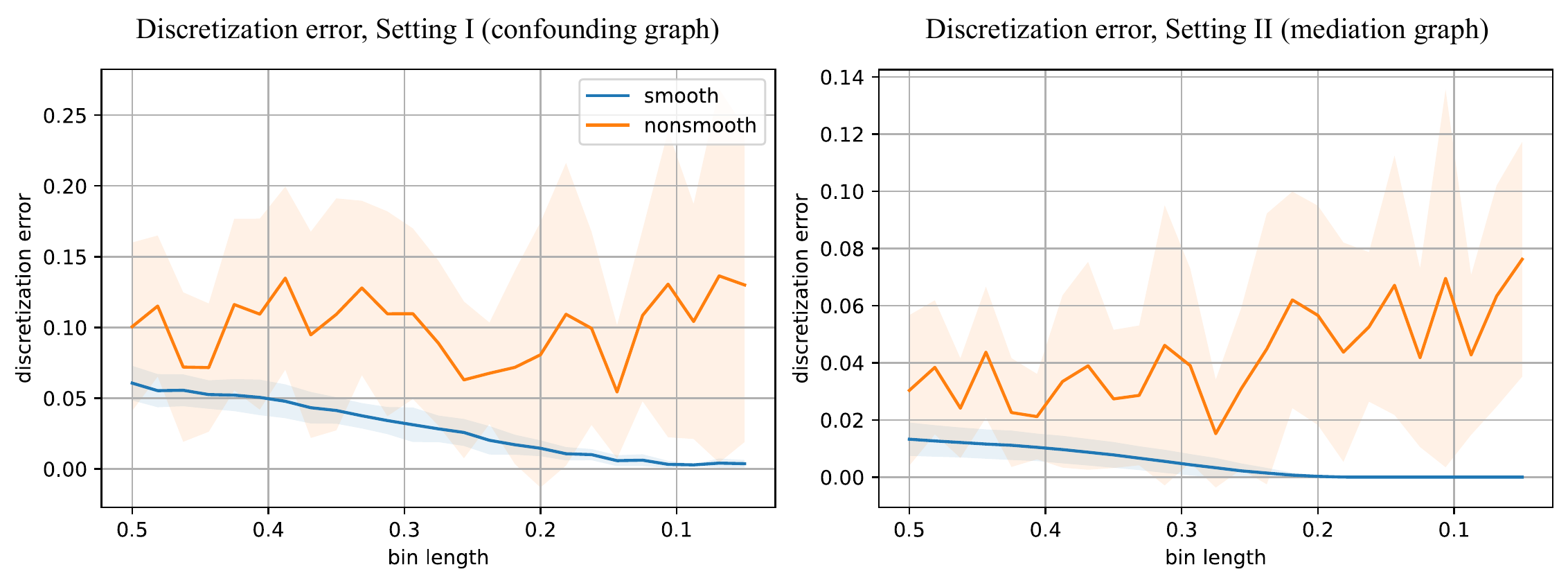}
    \caption{Discretization error with respect to the bin length. Left: setting I with the confounding graph Fig.~\ref{fig.proxy} (a). Right: setting II with the mediation graph Fig.~\ref{fig.proxy} (b). For both settings, the blue line corresponds to the case where the smoothness condition, \emph{i.e.}, Asm.~\ref{asm.tv-smooth-main}, holds, whereas the orange line corresponds to the case where the data is generated from a nonsmooth model.}
    \label{fig.dis-err-control}
\end{figure}

\textbf{Controlling the discretization error.} To demonstrate the effectiveness of our method in controlling the discretization error, we choose a pair of $(i,j)$ and compute the ${e}_{\mathrm{dis}}:=|{P}(\tilde{y}|\tilde{x}_i,\mathcal{U}_j)-{P}(\tilde{y}|\mathcal{U}_j)|$ with different bin lengths and present the results with blue lines in Fig.~\ref{fig.dis-err-control}. As we can see, when the smoothness condition in Asm.~\ref{asm.tv-smooth-main} is satisfied, the ${e}_{\mathrm{dis}}$ quickly decreases to zero as the bin length becomes small. This result aligns with our analysis in Prop.~\ref{prop.control-err-main}.

As a contrast, for the nonsmooth model (orange lines) in which Asm.~\ref{asm.tv-smooth-main} is violated, it is interesting to observe that the ${e}_{\mathrm{dis}}$ is consistently large for all bin lengths. This may be due to the large variation of the conditional distribution $\mathcal{L}_{Y|U=u}$ in the same discretization bin.

\begin{figure*}[htbp]
    \centering
    \includegraphics[width=0.9\textwidth]{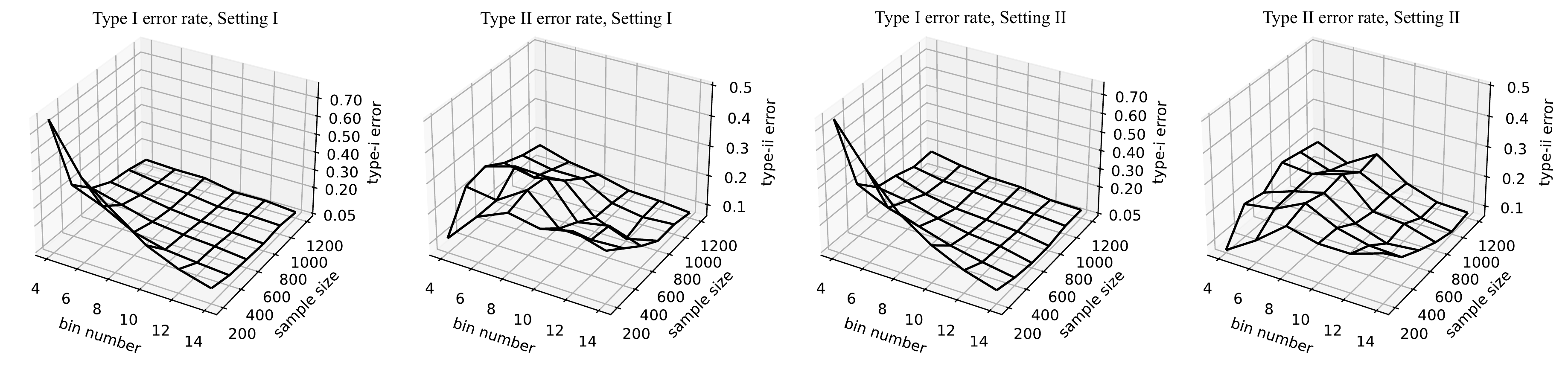}
    \caption{Type I and type II error rates with respect to the bin number and sample size. We consider two settings for data generating, with setting I (left) using the confounding graph in Fig.~\ref{fig.proxy} (a), and setting II (right) using the mediation graph in Fig.~\ref{fig.proxy} (b).}
    \label{fig.samplesize-binnumber}
\end{figure*}

\textbf{Influence of the bin number and the sample size.} In Fig.~\ref{fig.samplesize-binnumber}, we report the type I and type II error rates for different bin numbers $l_W=\{4,6,8,10,12,14\}$ (with $l_X = 14$ constantly) and sample sizes $n=\{200,400,600,800,1200\}$. As shown, as the bin number and the sample size grow, the type I error rate approaches the significance level $\alpha$, and the type II error rate becomes close to zero. These results align with the level and power of our testing procedure, as claimed in Thm.~\ref{thm.control.type-i} and Thm.~\ref{thm.type-ii-main}, respectively. Moreover, for small sample sizes (e.g., $n=200$), we can observe that the type II error rate slightly increases as the bin number grows, which may be due to insufficient samples in each bin to estimate the discrete probability.

\subsection{Application to sepsis disease}

\begin{figure}[htbp]
    \centering
    \includegraphics[width=.34\textwidth]{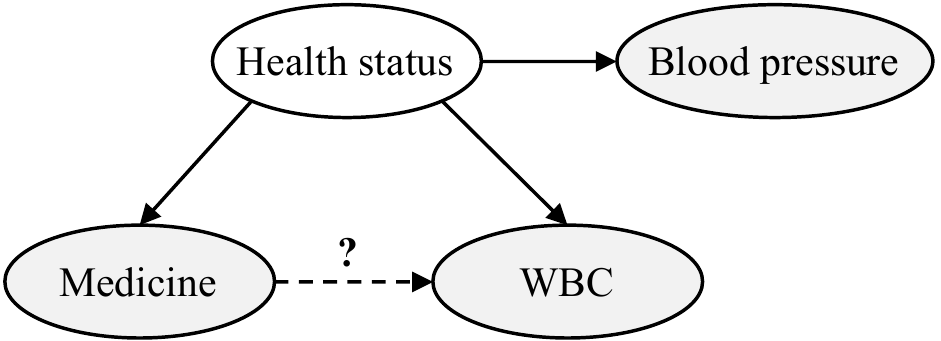}
    \caption{Illustration of causal discovery in sepsis disease. Observable variables are marked in gray. WBC denotes the count of White Blood Cells, which is a common biomarker used to assess patient's response to medicines. By using the blood pressure as the proxy variable ($W$) for the health status ($U$), our goal is to determine whether the edge $\mathrm{Medicine} \,-\!\!\to \mathrm{WBC}$ exists or not.}
    \label{fig.sepsis}
\end{figure}

In this section, we apply the proposed procedure to sepsis disease, which is a systemic inflammatory response to infection \cite{matot2001definition} and the leading cause of death in the intensive care unit (ICU) \cite{o2007sepsis}. 

Specifically, we examine two pairs of causal relationships: $\text{Vancomycin}\to \text{White Blood Cell count (WBC)}$ and $\text{Morphine}\to \text{WBC}$. Vancomycin and morphine are two common medications for sepsis, with vancomycin being used to control the bacterial infection \cite{moellering2006vancomycin} and morphine being used for pain management \cite{mcquay1999opioids}, and the WBC is a common biomarker used to evaluate patient's response to medications \cite{castelli2006procalcitonin}. 

In both pairs, it is believed that the confounder $U$ is accounted for by the patient's health status, which can influence the medication received and the WBC at the same time. To adjust for the latent confounding, we use blood pressure as the proxy variable ($W$). 


\begin{table}
\caption{Identified causal relationships in sepsis disease. We use the results from two random controlled trials (RCTs) \cite{anand2004effects,rosini2015randomized} as the golden standards. Results that contradict the golden standards are highlighted in red.}
\label{tab.sepsis}
\begin{center}
\resizebox{.42\textwidth}{!}{
\begin{tabular}{lccc}
\toprule
Method & Vancomycin$\to $WBC & Morphine$\to$ WBC \\
\midrule
RCT (gold. stand.)    & $\surd$ &  $\times$  \\
KCI & $\surd$ & \textcolor{red}{$\surd$}  \\
ICM  & $\surd$ & \textcolor{red}{$\surd$} \\
Miao    & \textcolor{red}{$\times$} &   \textcolor{red}{$\surd$} \\
Ours   & $\surd$ &  $\times$   \\
\bottomrule
\end{tabular}}
\end{center}
\vspace{-0.5cm}
\end{table}

\textbf{Golden standards from RCTs.} We take the results from two random controlled trials (RCTs) as the golden standards for causal discovery. Specifically, \cite{rosini2015randomized} found that vancomycin can causally influence the WBC; whereas \cite{anand2004effects} found that morphine has no causal effect on WBC. In other words, the correlation between morphine usage and an increase in WBC (as found in \cite{zhang2018prescription}) is likely due to the confounding bias.

\textbf{Data extraction from MIMIC database.} We consider the Medical Information Mart for Intensive Care (MIMIC III) database \cite{johnson2016mimic}, which contains the electronic health records for patients in the ICU. From MIMIC III, we extracted data for 3,251 patients diagnosed with sepsis during their ICU stays. Among these patients, 1,888 individuals received vancomycin, and 559 individuals received morphine.

\textbf{Results analysis.} In Tab.~\ref{tab.sepsis}, we report the identified causal relationships using our procedure and baseline methods. To highlight, we marked the results that are inconsistent with the result from RCT in red. As shown, in both pairs, our procedure identifies causal relations that well align with the RCT results, which demonstrate the effectiveness and utility of our method.

In contrast, there is a large discrepancy between the results of the baselines and the RCTs. For instance, KCI and ICM detect that morphine has a causal influence on WBC, which can be a false positive result due to the failure to adjust for the confounding bias; and Miao's procedure fails to reject the null hypothesis that vancomycin has no causal effect on WBC, which may be attributed to the loss of ability in the proxy variable without careful discretization.

\vspace{-0.105cm}
\section{Conclusions and Discussions}

In this paper, we propose a proxy-based hypothesis testing procedure for bivariate causal discovery. Our procedure involves the discretization of continuous variables, for which we control the discretization errors and establish the identifiability of a linear system. We then construct testing statistics based on the residue of linear regression, which enjoys asymptotically level and power.

\textbf{Limitations and future work.} Our method relies on discretization and hence may lose its power when the dimension of the conditional variable is high. Indeed, such a curse of dimensionality issue is common in all conditional independence (CI) tests \cite{ramdas2015decreasing}. For alleviation, we will investigate theories on high dimensional discretization (\citealt{van2014probability}, Lem.~5.12). Besides, we are also interested in plugging our test into multivariate causal discovery algorithms such as the FCI \cite{spirtes2000causation}, to enhance their ability for structure identification.

\section*{Impact statements} This paper presents work whose goal is to advance the fields of Causality and Machine Learning. There are many potential societal consequences of our work, none of which we feel must be specifically highlighted here.

\bibliography{reference}
\bibliographystyle{icml2024}

\newpage
\appendix
\onecolumn

\renewcommand{\contentsname}{Appendix}
\tableofcontents
\newpage
\addtocontents{toc}{\protect\setcounter{tocdepth}{2}}

\section{Notations \& definitions}

We introduce some extra notations and definitions that will be used in the proof. $f_X$ is the probability density function (pdf) of $X$, and similarly for the joint density $f_{X,Y}$ and conditional density $f_{X|Y}$. We say a pdf $f:\mathcal{X}\mapsto [0,1]$ is absolutely integrable if $\int_{\mathcal{X}} |f(x)| dx < \infty$. For clarity, we sometimes use the notations $\{\mathcal{X}_i\}_{i=1}^I,\{\mathcal{Y}_j\}_{j=1}^J,\{\mathcal{U}_k\}_{k=1}^K,\{\mathcal{W}_l\}_{l=1}^L$ to denote the partitions of $\mathcal{X},\mathcal{Y},\mathcal{U},$ and $\mathcal{W}$, respectively. For the matrix $A$, we use $\mathrm{det}(A)$ to denote its determinant. Besides, we use $\mathrm{ker}(A):=\{x:Ax=0\}$ and $\mathrm{im}(A):=\{y:y=Ax \,\, \text{for some} \,\, x\}$ to denote its kernel space and the image space, respectively.

\section{Discussion with works for causal direction identification}

There have been extensive studies in bivariate causal discovery. Within this area, one branch of these works distinguished cause from effect by exploiting asymmetry properties in functional causal models, which include but are not limited to the linear non-Gaussian model \cite{shimizu2006linear}, additive noise model \cite{hoyer2008nonlinear,hu2018causal,tu2022optimal}, and post non-linear model \cite{zhang2012identifiability,reick2021estimation}. Other methods are based on the independence of cause and mechanism \cite{sgouritsa2015inference,blobaum2018cause,tagasovska2020distinguishing}, and the stability of causation \cite{janzing2010causal,mitrovic2018causal,duong2022bivariate}. However, these works focused on learning causal direction, rather than establishing causal relations that can be more fundamental in real applications. Besides, the identifiability in these works relied on the strong ignorability condition, which may not hold in many scenarios. \textbf{In this paper}, we consider the causal relation identification problem in the presence of latent variables. 
\section{Discretization under completeness}

\subsection{Details of Exam.~\ref{exam.anm.completeness}: ANM with completeness}
\label{app-sec.example-comp}

\begin{example}(ANM when $U$ is a mediator)
    Suppose that $X,U,W$ satisfy the ANM, \emph{i.e.}, $U=g_U(X)+N_U, W=g_W(U)+N_W$, with $N_X,N_W$ being the noises. If $N_X$, $N_W$ both have non-zero characterization functions, then the conditional distributions $\mathcal{L}_{U|X}$ and $\mathcal{L}_{W|X}$ are complete.
    \label{exam.comp.mediator}
\end{example}

\begin{proof}
    We first show $\mathcal{L}_{U|X}$ is complete. Denote $h(x):=f_{N_U}(-x)$, for any bounded function $m$, we have 
    \begin{align*}
        \mathbb{E}[m(u)|x]=\int m(u)f_{N_U}(u-g_U(x))du = \int m(u)h(g_U(x)-u)du=\{m*h\} (g_U(x)),
    \end{align*}
     where $*$ denotes the convolution operator. Therefore, $\mathbb{E}[m(u)|x]=0$ a.s. means $m*h=0$ a.s. Applying the convolution theorem, we have $\hat{m} \cdot \hat{h}=0$ a.s., where $\hat{m},\hat{h}$ are the Fourier transforms of $m$ and $h$, respectively. Since the characterization function of $N_U$ is non-zero, we have $\hat{h}>0$ and hence $\hat{m}=0$ a.s. Applying the inverse Fourier transform, we have $m=0$ a.s., which means $\mathcal{L}_{U|X}$ is complete.

    Similarly, we can prove $\mathcal{L}_{W|U}$ is complete. Next, given that $\mathcal{L}_{U|X}$ and $\mathcal{L}_{W|U}$ are both complete, we show $\mathcal{L}_{W|X}$ is also complete. Specifically, for any bounded function $m$, we have 
    \begin{align*}
        \mathbb{E}[m(w)|x]=\int m(w)f_{W|X}(w|x) dw = \iint m(w)f_{W|U}(w|u)dw f_{U|X}(u|x) du = 0 \,\, \text{ a.s. }
    \end{align*}
    if and only if $\int m(w)f_{W|U}(w|u)dw = 0$ a.s., because $\mathcal{L}_{U|X}$ is complete. As $\mathcal{L}_{W|U}$ is also complete, we have $\int m(w)f_{W|U}(w|u)dw = 0$ a.s. if and only if $m(w)=0$ a.s. Therefore, we have $\mathbb{E}[m(w)|x]=0$ a.s. if and only if $m(w)=0$ a.s., which concludes the proof.
\end{proof}

\textbf{Example \ref{exam.anm.completeness}.} (ANM when $U$ is a confounder) Suppose that $X,U,W$ satisfy the ANM, \emph{i.e.}, $X=g_X(U)+N_X, W=g_W(U)+N_W$, with $N_X,N_W$ being the noises. If $g_X$ is invertible with a non-zero derivative, and $N_X$, $N_W$ have non-zero characterization functions, then the conditional distributions $\mathcal{L}_{U|X}$ and $\mathcal{L}_{W|X}$ are both complete.

\begin{proof}
    Similar to the proof in the Exam.~\ref{exam.comp.mediator}, we first have the completeness of $\mathcal{L}(W|U)$. Therefore, if we can prove the completeness for $\mathcal{L}(U|X)$, we can have the completeness of $E\{g(w)|x\}$, following the procedure in the mediation model. 

    To show $\mathcal{L}(W|U)$ is complete, note that we have $f_{U|X}(u|x) = f_{N_X}(x-g_{X}(u))|g_{X}^\prime(u)|$. Therefore, the completeness of $\mathcal{L}(U|X)$ means, for any bounded function $m$,
    \begin{align*}
        \int m(u)|g_{X}^\prime(u)| f_{N_X}(x-g_{X}(u))du = 0 \,\, \text{ a.s. if and only if } \,\, m(u) = 0 \,\, \text{ a.s. }
    \end{align*}
    Denote $m_1(u):=m(u)|g_X^\prime(u)|$. We first show that $m_1(u) = 0$ a.s. For this purpose, let $t=g_X(u)$ and hence $u=g_X^{-1}(t)$. Further, let $h(t)=g_X^{-1}(t)$ with $h^\prime(t)=\frac{1}{g_X^\prime(g_X^{-1}(t))}\neq 0$. Then, we have
    \begin{equation*}
        \int m_1(u) f_{N_X}(x-g_X(u)) du = \int m_1(h(t)) h^\prime(t) f_{N_X}(x-t) dt.
    \end{equation*}
    Since $N_X$ has a non-zero characteristic function, we first have $\int m_1(h(t)) h^\prime(t) f_{N_X}(x-t) dt = 0$ a.s. implies $m_1(h(t)) h^\prime(t)=0$ a.s. Since $h^\prime(t)\neq 0$, we then have $m_1(h(t))=m_1(g_X^{-1}(t))=0$, a.s. Since $u$ takes values in $\mathbb{R}$, we have the range of $g_X^{-1}$ is $\mathbb{R}$, and therefore $m_1(u)=0$, a.s. Since $g_X^\prime(u) \neq 0$, we have $m(u) = 0$ a.s., which concludes the proof.
\end{proof}

\subsection{Proof of Prop.~\ref{prop.rank-dis-main}: Rank-preserving discretization}
\label{app-sec.discrete}

\textbf{Proposition~\ref{prop.rank-dis-main}.} \emph{Suppose Asm.~\ref{asm.comp-main} holds. Then, for any discretization $\tilde{W}$ of $W$, there exists a discretization $\tilde{X}$ of $X$ such that the matrix $P(\tilde{W}|\tilde{X})$ has full row rank. Similarly, there also exists a discretization $\tilde{U}$ of $U$ such that the matrix $P(\tilde{W}|\tilde{U})$ is invertible.}

\begin{proof}
    We first show that for any partition $\{\mathcal{W}_l\}_{l=1}^L$, there exists a partition $\{\mathcal{X}_i\}_{i=1}^I$ (with $I\geq L$) such that the matrix ${P}(\tilde{W}|\tilde{X})$ has rank $L$. The proof contains two parts: in \textbf{part I}, we first show that for any $\{\mathcal{W}_l\}_{l=1}^L$, the functions $\{f_{W|X}(W\in\mathcal{W}_l|x)\}_{l=1}^L$ are linear independent. Then, in \textbf{part II}, we use this conclusion to construct $I$ points $\{x_i\}_{i=1}^I$ and therefore the partition $\{\mathcal{X}_i\}_{i=1}^I$ such that the matrix $P(\tilde{W}|\tilde{X})$ has rank $L$.

    \textbf{Part I}. We first show that for any $\{\mathcal{W}_l\}_{l=1}^L$, the functions $\{f_{W|X}(W\in\mathcal{W}_l|x)\}_{l=1}^L$ are linear independent. Prove by contradiction. Suppose that there exists a partition $\{\mathcal{W}_l\}_{l=1}^L$ such that $\{f_{W|X}(W\in\mathcal{W}_l|x)\}_{l=1}^L$ are linear dependent. That is, there are $L$ numbers $\{a_l\}_{l=1}^L$ that are not all zero such that $\sum_{l=1}^L a_l f_{W|X}(W\in\mathcal{W}_l|x) = 0$. Let $g(w)$ be the piece-wise constant function defined by $g(w):=a_l, w\in\mathcal{W}_l$, we then have $\int_{\mathcal{W}} g(w) f_{W|X}(w|x)=0$ for all $x$. Since $g(w)\neq 0$, this contradicts with the bounded completeness of $\mathcal{L}_{W|X}$ as assumed in Asm.~\ref{asm.comp-main}.

    We then show for any $\{\mathcal{W}_l\}_{l=1}^L$, the functions $\{F_l(x):=F_{X|W}(x|W\in\mathcal{W}_l)\}_{l=1}^L$ are linear independent. To see this, note that $f_{X|W}(x|W\in\mathcal{W}_l) = \frac{f_{W|X}(W\in\mathcal{W}_l|x)f_X(x)}{P(W\in\mathcal{W}_l)}$, therefore it is easy to obtain the linear independence of $\{f_{X|W}(x|W\in\mathcal{W}_l)\}_{l=1}^L$ through the independence of  $\{f_{W|X}(W\in\mathcal{W}_l|x)\}_{l=1}^L$. Besides, since $f_{X|W}(x|W\in\mathcal{W}_l)=F^\prime_{X|W}(x|W\in\mathcal{W}_l)$, we can also obtain the linear independence of $\{F_{X|W}(x|W\in\mathcal{W}_l)\}_{l=1}^L$.

    \textbf{Part II}. Given that $\{F_l(x)\}_{l=1}^L$ are linear independent, we now show there exists $I$ points $\{x_i\}_{i=1}^I$ such that the matrix $P(\tilde{W}|\tilde{X})$ has rank $L$. For this purpose, we only have to show that there exists $I$ points such that the $L$ vectors $F_1(\boldsymbol{x}),...,F_L(\boldsymbol{x})$ are linear independent, with $F_l(\boldsymbol{x}):=[F_l(x_1),...,F_l(x_I)]^T$ for each $l=1,..,L$. This is because, if this holds, we can complete the proof by showing that
    \begin{align*}
        \mathrm{rank}(P(\tilde{W}|\tilde{X}))=\mathrm{rank}(P(\tilde{X}|\tilde{W}))=\mathrm{rank}([F_1(\boldsymbol{x}),...,F_L(\boldsymbol{x})])=L.
    \end{align*}

    We first show the case where $I=L$. Denote the matrix $M_L=[F_1(\boldsymbol{x}),...,F_L(\boldsymbol{x})]$. We show $F_1(\boldsymbol{x}),...,F_L(\boldsymbol{x})$ are linear independent by showing $\mathrm{det}(M_L)\neq 0$. Prove by induction over $L$:
    
    (1) Base: when $L=1$, we have $\exists x_1, F_1(x_1)\neq0$. Therefore, we have $\mathrm{det}(M_1)=F_1(x_1)\neq 0$. 
    
    (2) Induction hypothesis: assume that for any $L-1$ linear independent functions $F_1,...,F_{L-1}$, there exists $L-1$ points $x_1,...,x_{L-1}$ such that $\mathrm{det}(M_{L-1})\neq 0$. 
    
    (3) Induction: show that for any $L$ linear independent functions $F_1,...,F_L$, there exists $L$ points $x_1,...,x_L$ such that $\mathrm{det}(M_L)\neq 0$. To this end, we first have the Laplace expansion of the matrix determinant 
    \begin{align}
    \label{eq.det}
        \mathrm{det}(M_L)(x_L)=\sum_{l=1}^L (-1)^{1+l} \mathrm{det}(A_l^L) F_l(x_L),
    \end{align}
    where $A_{L,l}$ denotes the matrix obtained by removing the $L$-th row and $l$-th column from $M_L$. That is, $A_{L,l}=[F_1(\boldsymbol{x}^\prime),...,F_{l-1}(\boldsymbol{x}^\prime),F_{l+1}(\boldsymbol{x}^\prime),...,F_L(\boldsymbol{x}^\prime)]$, where $F_l(\boldsymbol{x}^\prime)=[F_l(x_1),...,F_l(x_{L-1})]^T$. Then, since $\{F_1,...,F_{l-1},F_{l+1},...,F_L\}$ are $L-1$ linear independent functions, according to the induction hypothesis, there exists $L-1$ points $x_1,...,x_{L-1}$ such that $\mathrm{det}(A_{L,l})\neq 0$. That is, at least one of the cofactors $(-1)^{1+l} \mathrm{det}(A_l^L)$ in the expansion is not zero. Finally, since $F_1,...,F_L$ are linear independent, there exists $x_L$ such that $\mathrm{det}(M_L)(x_L) \neq 0$ for $x_L$, which concludes the induction.

    For cases where $I>L$, We have $F_1(\boldsymbol{x}),...,F_L(\boldsymbol{x})$ (where $F_l(\boldsymbol{x}):=[F_l(x_1),...,F_l(x_I)]^T$) are linear independent because $F_1(\boldsymbol{x}^\prime),...,F_L(\boldsymbol{x}^\prime)$ (where $F_l(\boldsymbol{x}^\prime):=[F_l(x_1),...,F_l(x_L)]^T$) are linear independent.

    Below, we show for any partition $\{\mathcal{W}_l\}_{l=1}^L$, there exists a partition $\{\mathcal{U}_k\}_{k=1}^K$ (with $K=L$) such that the matrix $P(\tilde{W}|\tilde{U})$ is invertible. We prove this by showing that under Assumption~\ref{asm.comp-main}, $\mathcal{L}_{W|U}$ is complete. Then, repeating the proof above, we have $P(\tilde{W}|\tilde{U})$ is invertible. To show this, according to the completeness of $\mathcal{L}_{W|X}$, for any bounded function $g$, we have
    \begin{align}
    \label{eq.complete-w-u}
        \int g(w)f(w|x)dw = \int \left(\int g(w)f(w|u)dw \right) f(u|x)du = 0 \text{ a.s. }
    \end{align}
    if and only if $g(w) = 0$ a.s. Moreover, this is also equivalent to $\int g(w) f(w|u)dw = 0$ a.s., since $\mathcal{L}_{U|X}$ is complete. Therefore, we have $\int g(w) f(w|u)dw = 0$ a.s. if and only if $g(w) = 0$ a.s., which concludes the proof.
\end{proof}
\section{Discretization error analysis}

\subsection{Definition of conditioning on a zero-probability event}
\label{sec.regular-prob}

The TV smoothness condition in Asm.~\ref{asm.tv-smooth-main} involves conditioning on the zero-probability event $U=u$, which may cause ambiguity without careful definition \cite{gyenis2017conditioning}. In the following, we use the regular conditional probability to provide a rigorous definition. We start from the standard Kolmogorov definition of conditional expectation.

\begin{definition}[Condition expectation, (\citealt{williams1991probability}, Def.~9.2)]
Let $(\Omega,\mathcal{F},P)$ be the probability space, $X$ a random variable with $E(|X|)<\infty$, and $\mathcal{G}$ a sub-$\sigma$-algebra of $\mathcal{F}$. Then there exists a random variable $Y$ such that:
\begin{enumerate}
    \itemsep0em
    \item $Y$ is $\mathcal{G}$-measurable,
    \item $E[|Y|]<\infty$,
    \item for every set $G\in \mathcal{G}$, we have:
    \begin{equation*}
        \int_G YdP = \int_G XdP.
    \end{equation*}
\end{enumerate}

Moreover, if $Y^\prime$ is another random variable with these properties then $Y^\prime=Y$, a.e., that is $P(\{w:Y(w)=Y^\prime(w)\})=1$. A random variable $Y$ with the above properties is called a version of the conditional expectation $E[X|\mathcal{G}]$ of $X$ given $\mathcal{G}$, and we write $Y=E[X|\mathcal{G}]$, a.s.
\end{definition}

\begin{remark}
    The existence of $Y$ satisfying the above properties is guaranteed by the Radon-Nikodym theorem (\citealt{durrett2019probability}, Thm. A.4.8). $Y=dX/d P$ is also called the Radon-Nikodym derivative.
\end{remark}

We often write $E[X|Z]$ in short for $E[X|\sigma(Z)]$, which is the conditional expectation of $X$ given $\sigma(Z)$. For $F\in \mathcal{F}$, define the conditional probability $P(F|Z)$ as a version of $E[\mathbbm{1}_{F}|Z]$. In this regard, the map $P(\cdot,\cdot): \Omega\times \mathcal{F}\mapsto [0,1)$ defined by $P(\omega,F):=P(F|Z)(\omega)$ is a measurable function that maps a tuple $(\omega,F)$ to a real value in $[0,1)$. 

Then, according to the disintegration theorem (\citealt{chang1997conditioning}, Def.~1 and Thm.~1), there exists a \textbf{regular conditional probability} $P(\cdot,\cdot): \Omega\times \mathcal{F}\mapsto [0,1)$ satisfying that:
\begin{enumerate}
    \itemsep0em
    \item for $F\in\mathcal{F}$, the function $\omega\mapsto P(\omega,F)$ is a version of $P(F|\mathcal{G})$,
    \item for almost every $\omega$, the map $F\mapsto P(\omega,F)$ is a probability measure on $\mathcal{F}$.
\end{enumerate}

\begin{remark}
    Note that for those $\{\omega\}$ where $P(\omega,F)$ is not well-defined (which is a set of measure zero), one can redefine $P(\omega,F):=\lim_{r\to 0} \frac{1}{|B_r(\omega)|}\int_{B_r(\omega)} P(t,F) dt$ (where $B_r(\omega)$ is a ball with radius $r$), so that $P(\omega,F)$ is now well-defined strictly everywhere.
\end{remark}

Equipped with the above definition of regular conditional probability (rcp), we then define the the conditional probability $P(X\in F|Z(\omega)=z)$ given the zero-probability event $Z=z$ as the rcp $P(\omega,F)$ (and similarly for the conditional probabilities $P(Y|U=u)$ and $P(W|U=u)$). Thanks to the above properties 1-2, our definition is well-defined.

\subsection{Details of Exam.~\ref{exam.tv-smooth}: ANM with TV smoothness}
\label{app-sec.tv-example}
\textbf{Example \ref{exam.tv-smooth}.} Suppose that $Y,U,W$ satisfy the ANM, \emph{i.e.}, $Y=g_Y(U)+N_Y, W=g_W(U)+N_W$, with $N_Y, N_W$ being the noises. If $g_Y$ and $g_W$ are differentiable, and the probability densities of $N_Y$ and $N_W$ have absolutely integrable derivatives, then the maps $u\mapsto \mathcal{L}_{Y|U=u}$ and $u\mapsto \mathcal{L}_{W|U=u}$ are Lipschitz continuous with respect to TV distance.

\begin{proof}
    We show $u\mapsto \mathcal{L}_{Y|U=u}$ is Lipschitz continuous and the Lipschitzness of $u\mapsto \mathcal{L}_{W|U=u}$ can be similarly derived. Specifically, according to the definition of TV distance, we have:
    \begin{align*}
        \mathrm{TV}(\mathcal{L}_{Y|U=u},\mathcal{L}_{Y|U=u^\prime}) & =\frac{1}{2} \int |f_{Y|U}(y|u) - f_{Y|U}(y|u^\prime)|dy, \\
        & \overset{(1)}{=} \frac{1}{2} \int |f_{N_Y}(y-g_Y(u)) - f_{N_Y}(y-g_Y(u^\prime))|dy, \\
        & \overset{(2)}{=} \int \frac{1}{2} |f_{N_Y}^\prime(y-g_Y(u_c)) \cdot g_Y^\prime(u_c)| \cdot |u-u^\prime| dy.
    \end{align*}
    for some $u_c \in (u,u^\prime)$, where ``(1)" is due to ANM that can write $f_{Y|U}(y|u)$ as $f_{N_Y}(y-g_Y(u))$ and ``(2)" is due to the mean value theorem. Let $L_Y:= \frac{1}{2}\int |f_{N_Y}^\prime(y-g_Y(u_c)) g_Y^\prime(u_c)| dy$, we have $\mathrm{TV}(\mathcal{L}_{Y|U=u},\mathcal{L}_{Y|U=u^\prime})\leq L_Y |u-u^\prime|$, which concludes the proof.
\end{proof}

\subsection{Proof of Prop.~\ref{prop.control-err-main}: Discretization error control under smoothness}

\textbf{Proposition~\ref{prop.control-err-main}.} \emph{Suppose that $\{\mathcal{U}_k\}_{k=1}^K$ is a partition of $\mathcal{U}$, and that for every bin $\mathcal{U}_k$,
\begin{equation*}
    len(\mathcal{U}_k)\leq \frac{1}{2}min\{\epsilon/L_Y,\epsilon/L_W\}.
\end{equation*}
Then, under Asm.~\ref{asm.tv-smooth-main} and $\mathbb{H}_0:X\ind Y|U$, we have:
\begin{align*}
    |p(\tilde{y}|\tilde{u}_k) - p(\tilde{y}|\tilde{x},\tilde{u}_k)| \leq \epsilon, \quad |p(\tilde{w}|\tilde{u}_k) - p(\tilde{w}|\tilde{x},\tilde{u}_k)| \leq \epsilon.
\end{align*}}

\begin{proof}
    First recall that we use the notations $\{\mathcal{X}_i\}_{i=1}^I,\{\mathcal{Y}_j\}_{j=1}^J,\{\mathcal{U}_k\}_{k=1}^K,\{\mathcal{W}_l\}_{l=1}^L$ to denote the partitions of $\mathcal{X},\mathcal{Y},\mathcal{U},$ and $\mathcal{W}$, respectively. Below, we show $|P(Y\in\mathcal{Y}_j|X\in\mathcal{X}_i,U\in\mathcal{U}_k) - P(Y\in\mathcal{Y}_j|U\in\mathcal{U}_k)|\leq \epsilon$ and the bounding of $|P(W\in\mathcal{W}_l|X\in\mathcal{X}_i,U\in\mathcal{U}_k) - P(W\in\mathcal{W}_l|U\in\mathcal{U}_k)|$ can be similarly derived. 
    
    The proof consists of two parts. In \textbf{part I}, we show the smoothness of $u\mapsto\mathcal{L}_{Y|U=u}$, which together with $\mathrm{len}(\mathcal{U}_k)\leq \epsilon/(2L_Y)$, implies $\mathrm{TV}(\mathcal{L}_{Y|U=u},\mathcal{L}_{Y|U\in\mathcal{U}_k})\leq\epsilon/2$. Then, in \textbf{part II}, we show the control $|P(Y\in\mathcal{Y}_j|X\in\mathcal{X}_i,U\in\mathcal{U}_k) - P(Y\in\mathcal{Y}_j|U\in\mathcal{U}_k)|$ via that of $\mathrm{TV}(\mathcal{L}_{Y|U=u},\mathcal{L}_{Y|U\in\mathcal{U}_k})$.

    \textbf{Part I}. We show $\mathrm{TV}(\mathcal{L}_{Y|U=u},\mathcal{L}_{Y|U\in\mathcal{U}_k})\leq\epsilon/2$. Note that we first have 
    \begin{align*}
        \mathcal{L}_{Y|U\in\mathcal{U}_k} = \dashint_{\mathcal{U}_k} \mathcal{L}_{Y|U=u} d\mathcal{L}_{U}(u) := \frac{\int_{\mathcal{U}_k} \mathcal{L}_{Y|U=u} d\mathcal{L}_{U}(u)}{\int_{\mathcal{U}_k} d\mathcal{L}_{U}(u)},
    \end{align*}
    where $\dashint$ denotes the average integral \cite{warren2021wasserstein}. To see this, according to the definition of $\mathcal{L}$, for any $\mathcal{Y}_j\in\mathcal{Y}$, the LHS equals $\mathcal{L}_{Y|U\in\mathcal{U}_k}(\mathcal{Y}_j) = P(Y\in\mathcal{Y}_j|U\in\mathcal{U}_k)$, and the RHS equals ${\int_{\mathcal{U}_k} \mathcal{L}_{Y|U=u}(\mathcal{Y}_j) d\mathcal{L}_{U}(u)}/{P(U\in\mathcal{U}_k)} = {P(Y\in\mathcal{Y}_j,U\in\mathcal{U}_k)}/{P(U\in\mathcal{U}_k)}=\mathrm{LHS}$.

    Then, due to the assumed TV smoothness, for any $u,u^\prime\in \mathcal{U}_K$, we have $\mathrm{TV}(\mathcal{L}_{Y|U=u},\mathcal{L}_{Y|U=u^\prime}) := \sup_{\mathcal{Y}_j\in \mathcal{Y}} |\mathcal{L}_{Y|U=u}(\mathcal{Y}_j)-\mathcal{L}_{Y|U=u^\prime}(\mathcal{Y}_j)| \leq L_Y|u-u^\prime|\leq \epsilon/2$. Now, for \textbf{\emph{any}} $\mathcal{Y}_j \subseteq \mathcal{Y}$, we have (note that $u$ is a fixed point and $u^\prime$ varies in the integration domain):
    \begin{align*}
        |\mathcal{L}_{Y|U=u}(\mathcal{Y}_j) &- \mathcal{L}_{Y|U\in \mathcal{U}_k}(\mathcal{Y}_j)| = \left| \mathcal{L}_{Y|U=u}(\mathcal{Y}_j) - \dashint_{\mathcal{U}_k} \mathcal{L}_{Y|U=u^\prime}(\mathcal{Y}_j) d\mathcal{L}_U(u^\prime)\right| \\
        &=\left| \dashint_{\mathcal{U}_k} \mathcal{L}_{Y|U=u}(\mathcal{Y}_j) d\mathcal{L}_U(u^\prime) - \dashint_{\mathcal{U}_k} \mathcal{L}_{Y|U=u^\prime}(\mathcal{Y}_j) d\mathcal{L}_U(u^\prime)\right| && \text{\#} \dashint_{\mathcal{U}_k} d\mathcal{L}_U(u^\prime) = 1 \\
        &\leq \dashint_{\mathcal{U}_k} \left| \mathcal{L}_{Y|U=u}(\mathcal{Y}_j) - \mathcal{L}_{Y|U=u^\prime}(\mathcal{Y}_j)\right| d\mathcal{L}_U(u^\prime) && \text{\# Triangle inequity} \\
        &\leq \dashint_{\mathcal{U}_k} \epsilon/2 \, d\mathcal{L}_U(u^\prime) = \epsilon/2, 
    \end{align*}
    which means $\sup_{\mathcal{Y}_j\in\mathcal{Y}} |\mathcal{L}_{Y|U=u}(\mathcal{Y}_j)-\mathcal{L}_{Y|U\in\mathcal{U}_k}(\mathcal{Y}_j)| =: \mathrm{TV}(\mathcal{L}_{Y|U=u},\mathcal{L}_{Y|U\in\mathcal{U}_k}) \leq \epsilon/2$. In a similar way, we can prove $\mathrm{TV}(\mathcal{L}_{Y|X\in\mathcal{X}_i,U=u},\mathcal{L}_{Y|X\in\mathcal{X}_i,U\in\mathcal{U}_k})\leq\epsilon/2$. To be specific, under $\mathbb{H}_0:X\ind Y|U$, we have $\mathcal{L}_{Y|X\in\mathcal{X}_i,U=u}=\mathcal{L}_{Y|U=u}$ for any $\mathcal{X}_i\subseteq\mathcal{X}$. As a result, the map $u\mapsto \mathcal{L}_{Y|X\in\mathcal{X}_i,U=u}$ is also $L_Y$-Lipschitz continuous wrt TV distance. The rest of the proof is the same as above. 

    \textbf{Part II.} We show $|P(Y\in\mathcal{Y}_j|X\in\mathcal{X}_i,U\in\mathcal{U}_k) - P(Y\in\mathcal{Y}_j|U\in\mathcal{U}_k)|\leq \epsilon$. Specifically, since $\mathrm{TV}(\mathcal{L}_{Y|U=u},\mathcal{L}_{Y|U\in\mathcal{U}_k}) \leq \epsilon/2$, we have $|P(Y\in\mathcal{Y}_j|U=u) - P(Y\in\mathcal{Y}_j|U\in\mathcal{U}_k)|\leq \epsilon/2$. Since $\mathrm{TV}(\mathcal{L}_{Y|X\in\mathcal{X}_i,U=u},\mathcal{L}_{Y|X\in\mathcal{X}_i,U\in\mathcal{U}_k})\leq \epsilon/2$, we also have $|P(Y\in\mathcal{Y}_j|X\in\mathcal{X}_i,U=u) - P(Y\in\mathcal{Y}_j|X\in\mathcal{X}_i,U\in\mathcal{U}_k)|\leq \epsilon/2$. According to the triangle inequality, we have 
    \begin{align*}
        & |P(Y\in\mathcal{Y}_j|U\in\mathcal{U}_k) - P(Y\in\mathcal{Y}_j|X\in\mathcal{X}_i,U\in\mathcal{U}_k)| \\
        \leq & |P(Y\in\mathcal{Y}_j|U\in\mathcal{U}_k) - P(Y\in\mathcal{Y}_j|U=u)| + |P(Y\in\mathcal{Y}_j|U=u) - P(Y\in\mathcal{Y}_j|X\in\mathcal{X}_i,U\in\mathcal{U}_k)|\\
        = & |P(Y\in\mathcal{Y}_j|U\in\mathcal{U}_k) - P(Y\in\mathcal{Y}_j|U=u)| + |P(Y\in\mathcal{Y}_j|X\in \mathcal{X}_i, U=u) - P(Y\in\mathcal{Y}_j|X\in\mathcal{X}_i,U\in\mathcal{U}_k)| \leq \frac{\varepsilon}{2} + \frac{\varepsilon}{2} = \varepsilon,
    \end{align*}
    where the 2nd equality is due to $P(Y\in\mathcal{Y}_j|U=u)=P(Y\in\mathcal{Y}_j|X\in\mathcal{X}_i,U=u)$ under $\mathbb{H}_0:X\ind Y|U$.
\end{proof}

\subsection{Details of Exam.~\ref{exam.anm.tight}: ANM with tightness}

\textbf{Example \ref{exam.anm.tight}.} (ANM when $U$ is a confounder) Suppose that $X$ and $U$ satisfy the ANM, \emph{i.e.}, $X=g_X(U)+N_X$, with $N_X$ being the noise. If $g_X$ is invertible and the distribution of $N_X$ is tight, then $\mathcal{L}_{U|X}$ is uniformly tight in an arbitrary compact interval $[a,b]$.

\begin{proof}
    Since $\mathcal{L}_{N_U}$ is tight, we first have $\forall \epsilon>0, \exists t_0, \forall t\geq t_0, \int_t^\infty f_{N_U}(u)du \leq \epsilon$. Then we can obtain the uniform tightness of $\mathcal{L}_{U|X}$ by noting that $\forall \epsilon > 0$, we have $s_0:=g_X^{-1}(a-t_0)$ such that $\forall s \geq s_0$, 
    \begin{align*}
            \int_{-\infty}^t f_{U|X}(u|x)du &= - \int_{-\infty}^t f_{N_U}(x-g_X(u)) d(x-g_X(u)) \\
            & = \int_{x-g_X(t)}^{x-g_X(-\infty)} f_{N_U}(n) dn \leq  \int_{x-g_X(t)}^{\infty} f_{N_U}(n) dn \leq \epsilon,
        \end{align*}
        where we use the substitution $n=x-g_X(u)$, and the fact that $x-g_X(t)\geq x-a+t_0\geq t_0$.
\end{proof}

\begin{example}[ANM when $U$ is a mediator]
    Suppose that $X$ and $U$ satisfy the ANM, \emph{i.e.}, $U=g_U(X)+N_U$, with $N_U$ being the noise. If $g_U$ is bounded in $[a,b]$ and the distribution of $N_U$ is tight, then $\mathcal{L}_{U|X}$ is uniformly tight in $[a,b]$.
\end{example}

\begin{proof}
    Since $\mathcal{L}_{N_U}$ is tight, we first have $\forall \epsilon>0, \exists t_0, \forall t\geq t_0, \int_t^\infty f_{N_U}(u)du \leq \epsilon$. Besides, we denote $m_0$ as the upper bound of $g_U(x)$, such that $|g_U(x)| \leq m_0$ for all $x \in [a,b]$. Then we can obtain the uniform tightness of $\mathcal{L}_{U|X}$ by noting that $\forall \epsilon > 0$, we have $s_0:=t_0+m_0$ such that $\forall s \geq s_0$, 
    \begin{align*}
        \int_{s}^\infty f_{U|X}(u|x) du = \int_s^\infty f_{N_U}(u-g_X(x)) du = \int_{s-g_U(x)}^\infty f_{N_U}(u) du \leq \epsilon, 
    \end{align*}
    where we use the fact that $s\geq t_0 + h_0 \geq t_0 + |g_U(x)| \geq t_0 + g_U(x)$ therefore  $s-g_U(x)\geq t_0$.
\end{proof}

\subsection{Proof of Prop.~\ref{prop.dis-err unbd}: Discretization error control under tightness}

\textbf{Proposition~\ref{prop.dis-err unbd}.} \emph{Suppose that $\mathcal{L}_{U|X}$ is uniformly tight in $[a,b]$, and that the map $x\mapsto\mathcal{L}_{U|X=x}$ is $L_U$-Lipschitz continuous wrt TV distance. Suppose that $\{\mathcal{X}_i\}_{l_X}$ is a partition of $[a,b]$ satisfying $len(\mathcal{X}_i)\leq \epsilon/(2L_U)$. Then, $\forall\epsilon>0, \exists t_0$, let $\mathcal{U}_{nc}:=\{u:\Vert U \Vert_2 > t_0\}$, we have $p(\tilde{u}_{nc}|\tilde{x}_i) \leq \epsilon$.}

\begin{proof}
    We only show the proof of $P(Y\in \mathcal{Y}_j,\Vert U \Vert_2 > t_0|X\in\mathcal{X}_i)\leq \epsilon$, as the control of $P(W\in \mathcal{W}_l,\Vert U \Vert_2 > t_0|X\in\mathcal{X}_i)$ can be similarly derived. We obtain this conclusion by showing $P(\Vert U \Vert_2 > t_0|X\in\mathcal{X}_i)\leq \epsilon$ and using the fact that
    \begin{align*}
        P(Y\in \mathcal{Y}_j,\Vert U \Vert_2 > t_0|X\in\mathcal{X}_i) = P(Y\in \mathcal{Y}_j|\Vert U \Vert_2 > t_0,X\in\mathcal{X}_i) P(\Vert U \Vert_2 > t_0|X\in\mathcal{X}_i) \leq P(\Vert U \Vert_2 > t_0|X\in\mathcal{X}_i)\leq \epsilon.
    \end{align*}
   Specifically, given that $\mathcal{L}_{U|X}$ is uniformly tight, we first have $\forall \epsilon/2>0, \exists t_0, \forall t\geq t_0, \forall x\in \mathcal{X}_i, P(\Vert U \Vert_2 > t|x) \leq \epsilon/2$. Then, since $x\mapsto\mathcal{L}_{U|X=x}$ is $L_U$-Lipschitz, we have $\forall \epsilon/2>0, |P(|U|>t|x) - P(|U|>t|\mathcal{X}_i)|\leq \epsilon/2$ for $x\in\mathcal{X}_i$. Applying the triangle inequality, we have $P(|U|>t|\mathcal{X}_i)\leq \epsilon$, which concludes the proof.
\end{proof}
\section{Hypothesis-testing procedure}

\subsection{Maximum likelihood estimation}
\label{app-sec.mle}

Consider discrete random variables $X,Y$ that respectively take values $\{x_1,...,x_I\}$ and $\{y_1,...,y_J\}$. Suppose that we have $n$ iid samples. Denote $Z_{ij}$ the number of samples that take the value $(x_i,y_j)$, and denote $Z_{i:}:=\sum_j Z_{ij}$. Then, the MLE estimator of conditional probability vector $P(y|X)$ is $\hat{P}(y|X) = \left[\hat{p}(y|x_1),...,\hat{p}(y|x_I)\right]$, where $\hat{p}(y_j|x_i) = Z_{ij}/Z_{i:}$.

The MLE estimator satisfies that $\hat{P}(y|X)\to P(y|X)$ in probability, and that $n^{1/2}(\hat{P}(y|X)-P(y|X)) \to N(0,\Sigma)$ in distribution. The covariance matrix is $\Sigma=\mathrm{diag}[\frac{p(y_j|x_1)(1-p(y_j|x_1))}
{p(x_1)},...,\frac{p(y_j|x_I)(1-p(y_j|x_I))}{p(x_I)}]$ and is positive definite. The corresponding estimator $\hat{\Sigma}$ that satisfies $\hat{\Sigma}\to \Sigma$ in probability can be obtained by replacing $p(y_j|x_i)$ with $\hat{p}(y_j|x_i)$, and replacing $p(x_i)$ with $\hat{p}(x_i)=Z_{i:}/n$. These results are obtained by noting that $Z_{ij}$ follows the multinomial distribution, and by applying the Fisher information matrix (in the sandwich form) and the Cramer-Rao lower bound (\citealt{casella2021statistical}, Thm.~7.3.9).

\subsection{Proof of Thm.~\ref{thm.control.type-i}: Asymptotic level of the test}

\textbf{Theorem \ref{thm.control.type-i}.} \emph{Suppose that Asm.~\ref{asm.comp-main},~\ref{asm.tv-smooth-main},~\ref{asm.tight-main} hold, then under $H_0$, it holds that:
    \begin{equation*}
        T_{n,l} \to \chi_{r}^2 \,\, \text{in distribution}
    \end{equation*}
    as $n,l\to \infty$. As a result, for any $\alpha\in (0,1)$, we have:
    \begin{equation*}
        \lim_{n,l\to\infty} P(T_{n,l}> z_{1-\alpha}) = \alpha.
    \end{equation*}
    In other words, the testing procedure we define has uniform asymptotic level $\alpha$.}

\begin{proof}
    First recall that we use the notations $\{\mathcal{X}_i\}_{i=1}^I,\{\mathcal{Y}_j\}_{j=1}^J,\{\mathcal{U}_k\}_{k=1}^K,\{\mathcal{W}_l\}_{l=1}^L$ to denote the partitions of $\mathcal{X},\mathcal{Y},\mathcal{U},$ and $\mathcal{W}$, respectively. We use $l$ and $L$ interchangeably, and denote $r:=I-L$. We denote $q:=P(\tilde{y}|\tilde{X})^{T}$ and $Q:=P(\tilde{W}|\tilde{X})$.
    
    The proof consists of two parts. In \textbf{part I}, we first show the asymptotic linearity between $q^T$ and $Q$. Then, in \textbf{part II}, we analyze the asymptotic distribution of the test statistics.

    \textbf{Part I.} In the following, we show the asymptotic linearity $q^T=P(\tilde{y}|\tilde{U})P(\tilde{W}|\tilde{U})^{-1}Q + \boldsymbol{e}$ with $\lim_{L\to\infty} \boldsymbol{e}=\boldsymbol{0}$.
    
    We first consider cases where $\mathcal{U}$ is a compact interval. To show the asymptotic linearity , we first show:
    \begin{align}
        q^T & = P(\tilde{y}|\tilde{U})P(\tilde{U}|\tilde{X}) + \boldsymbol{e}_1, \ \lim_{L\to\infty} \boldsymbol{e}_1 = \boldsymbol{0}, \label{eq.linear-q}\\
        Q & = P(\tilde{W}|\tilde{U})P(\tilde{U}|\tilde{X}) + \boldsymbol{e}_2, \ 
        \lim_{L\to\infty} \boldsymbol{e}_2 = \boldsymbol{0}. \label{eq.linear-Q}
    \end{align}
    In this regard, under Asm.~\ref{asm.comp-main}, we have: 
    \begin{align*}
        q^T= P(\tilde{y}|\tilde{U})P(\tilde{W}|\tilde{U})^{-1} Q + \boldsymbol{e}, \ \lim_{L\to\infty}\boldsymbol{e}=\lim_{L\to\infty} \{\boldsymbol{e}_1 - P(\tilde{y}|\tilde{U})P(\tilde{W}|\tilde{U})^{-1} \boldsymbol{e}_2\} =\boldsymbol{0},
    \end{align*}
    which concludes \textbf{part I}. To show Eq.~\eqref{eq.linear-q}, note that we first have 
    \begin{align*}
        q^T:=P(\tilde{y}|\tilde{X}) = P(\tilde{y}_l|\tilde{U})P(\tilde{U}|\tilde{X}) + \boldsymbol{e}_1, \ \boldsymbol{e}_1 = [\{P(\tilde{y}_l|\tilde{U},\tilde{x}_i) - P(\tilde{y}_l|\tilde{U})\}P(\tilde{U}|\tilde{x}_i)]_{i=1}^I.
    \end{align*}
    Then, according to Prop.~\ref{prop.control-err-main} and the mean inequality, we have $\lim_{L\to\infty} \boldsymbol{e}_1 = \boldsymbol{0}$.Similarly, we can show
    \begin{align*}
        Q = P(\tilde{W}|\tilde{U})P(\tilde{U}|\tilde{X}) + \boldsymbol{e}_2, \text{ where } \boldsymbol{e}_2 = [\{P(\tilde{W}|\tilde{U},\tilde{x}_i) - P(\tilde{W}|\tilde{U})\}P(\tilde{U}|\tilde{x}_i)]_{i=1}^I, 
    \end{align*}
    and $\lim_{L\to\infty} \boldsymbol{e}_2 = \boldsymbol{0}$.

    We then consider the case where $\mathcal{U}=\mathbb{R}$. Still, we want to establish Eq.~\eqref{eq.linear-q}-~\eqref{eq.linear-Q}. We first derive the form of $\boldsymbol{e}_1$. For a given $t$, divide $\mathcal{U}$ into $\mathcal{U}^\prime:=\{u:|u|>t\}$ and $\mathcal{U}^{''}:=\mathcal{U}\backslash\mathcal{U}^\prime$, and further partition $\mathcal{U}^{''}$ into $K$ bins $\{\mathcal{U}_k\}_{k=1}^K$. We then have 
    \begin{align*}
        p(\tilde{y}|\tilde{x}_i) = P(\tilde{y}|\tilde{U},\tilde{x}_i) P(\tilde{U}|\tilde{x}_i) + p(\tilde{y}|U\in\mathcal{U}^\prime,\tilde{x}_i) p(U\in\mathcal{U}^\prime|\tilde{x}_i).
    \end{align*}
    Therefore, we have: 
    \begin{align*}
        P(\tilde{y}|\tilde{X}) & = P(\tilde{y}|\tilde{U})P(\tilde{U}|\tilde{X}) + \boldsymbol{e}_{\text{c}} + \boldsymbol{e}_{\text{nc}}, \\
        \boldsymbol{e}_{\text{c}} & = \{[P(\tilde{y}|\tilde{U},\tilde{x}_i)-P(\tilde{y}|\tilde{U})] P(\tilde{U}|\tilde{x_i})\}_{i=1}^I, \\
        \boldsymbol{e}_{\text{nc}} & = \{p(\tilde{y},U\in\mathcal{U}^\prime|\tilde{x}_i)p(U\in\mathcal{U}^\prime|\tilde{x}_i)\}_{i=1}^I. 
    \end{align*}
    where $\boldsymbol{e}_{\text{c}}$ is the error in the compact interval $\mathcal{U}^\prime$ and $\boldsymbol{e}_{\text{nc}}$ is the error in the non-compact $\mathcal{U}^{''}$. We then have $\lim_{L \to \infty} \boldsymbol{e}_1=\lim_{L \to \infty}\boldsymbol{e}_{\text{c}} + \boldsymbol{e}_{\text{nc}} = \boldsymbol{0}$, since $\lim_{L \to \infty} \boldsymbol{e}_{\text{c}} = \boldsymbol{0}$ by repeating the previous proof and $\lim_{L \to \infty} \boldsymbol{e}_{\text{nc}} = \boldsymbol{0}$ can be shown in Prop.~\ref{prop.dis-err unbd}. Similarly, we have:
    \begin{align*}
        Q=P(\tilde{W}|\tilde{U})P(\tilde{U}|\tilde{X}) + \boldsymbol{e}_{\text{2}}, \ \boldsymbol{e}_2=\{[P(\tilde{W}|\tilde{U},\tilde{x}_i)-P(\tilde{W}|\tilde{U})] P(\tilde{U}|\tilde{x_i})\}_{i=1}^I+\{P(\tilde{W}|U\in\mathcal{U}^\prime,\tilde{x}_i)P(U\in\mathcal{U}^\prime|\tilde{x}_i)\}_{i=1}^I, 
    \end{align*}
    and $\lim_{t,L\to\infty} \boldsymbol{x}_2=\boldsymbol{0}$.

    \textbf{Part II.} We analyze the asymptotic distribution of the statistics $T:=n^{1/2}\xi^T \xi$. We first show $n^{1/2} \xi \to N(0,\Omega)$. Specifically, given that $\hat{Q} \to Q$, $\hat{\Sigma} \to \Sigma$ in probability, and that $n^{1/2}(\hat{q} - q) \to N(0,\Sigma)$ in distribution, apply Slutsky's theorem, we have: 
    \begin{align}
    \label{eq.omega}
        n^{1/2}(\xi - \Omega\Sigma^{1/2}q_l)\to_d N(0,\Omega_l), \ \Omega:=\mathrm{I}_I - {\Sigma}^{-1/2} {Q}^T ({Q} {\Sigma}^{-1} {Q}^T)^{-1} {Q} {\Sigma}^{-1/2}.
    \end{align}
    Then, under $\mathbb{H}_0:X\ind Y|U$, according to the asymptotic linearity in \textbf{part I}, we have $\Omega\Sigma^{-1/2}q \to 0$ as $L\to\infty$. Applying Slutsky's theorem again, we have $n^{1/2} \xi \to N(0,\Omega)$ in distribution. 
    
    Then, we show $\mathrm{rank}(\Omega)=r$. Specifically, we have $\mathrm{rank}(Q)=L$ according to Prop.~\ref{prop.rank-dis-main}. Since $\Sigma^{-1/2}$ is invertible, we have $\mathrm{rank}(Q\Sigma^{-1/2})=L$. Thus, the idempotent matrix $M:={\Sigma}^{-1/2} {Q}^T ({Q} {\Sigma}^{-1} {Q}^T)^{-1} {Q} {\Sigma}_l^{-1/2}$ has rank $L$. According to Coro.~11.5 of \citep{banerjee2014linear}, $M$ has $K$ eigenvalues equal to one and $I-L$ eigenvalues equal to zero. Thus, we have $\Omega=\mathrm{I}-M$ ($\mathrm{I}$ denotes the identity matrix) being an idempotent matrix with rank $r=I-L$.

    Finally, we show $T \to \chi_{r}^2$ in distribution. To be specific, given that $n^{1/2} \xi \to N(0,\Omega)$, we have $T:=n\xi^T \xi \to N(0,\Omega)^T N(0,\Omega)$ according to Thm.~1.12 of \cite{shao2003mathematical}. What is left is to show $N(0,\Omega)^T N(0,\Omega)=\chi^2_r$. To see this, note that we have the orthogonal decomposition $V\Omega V^T= \mathrm{diag}(1,...,1,0,...,0)$, where $V$ is an orthogonal matrix and $\mathrm{diag}(1,...,1,0,...,0)$ is a diagonal matrix with $r$ $1$s and $(I-r)$ $0$s in the diagonal. Therefore, we have $VN(0,\Omega) = N(0,\mathrm{diag}(1,...,1,0,...,0))$ and $N(0,\Omega)^T N(0,\Omega) = \left[VN(0,\Omega)\right]^T \left[VN(0,\Omega)\right] = \chi^2_r$.
\end{proof}

\subsection{Details of Rem.~\ref{rem.type-ii-cond}: Characterization of Asm.~\ref{asm.int-eq-main}}
\label{app-sec.int-eq} 

The asymptotic power of our test (Thm.~\ref{thm.type-ii-main}) is built on a local alternative $\mathbb{H}_1$ that the integral equation $f(y|x)=\int h(y,w)f(w|x)dw$ does not have a solution. Intuitively, $\mathbb{H}_1$ means that the correlation between $X$ and $Y$ can not be fully explained by that between $X$ and $W$. Empirically, we find this condition can hold in many cases under $X\not\ind Y|U$. Specifically, we conduct numerical experiments with the following models:

\begin{itemize}
\itemsep0em
    \item {Linear Gaussian model:} $U=X+N_1, Y=a X+U+N_2, W=U+N_3, \  X, N_1, N_2, N_3 \sim N(0, 0.1)$, where $a$ ranges from $1$ to $10$.
    \item {Linear non-Gaussian model:} $U=X+N_1, Y=a X+U+N_2, W=U+N_3,  X \sim \mathrm{ Uniform}(-0.5, 0.5), N_1, N_2, N_3 \sim \operatorname{Exp}(0.1)$, where $a$ ranges from $1$ to $10$. 
    \item {ANM:} $U=X^2+N_1, Y=a e^X+U+N_2,  W=\tan (U)+N_3, \ X \sim \mathrm{Uniform}(-0.5, 0.5), \ N_1, N_2, N_3 \sim \operatorname{Exp}(0.1)$, where $a$ ranges from $1$ to $10$. 
    \item {General nonlinear model:} $U=|X| N_1$, $Y=a e^{X+N_2}+U$, \  $W=\tan \left(U+N_3\right)$, $\ X \sim \mathrm{Uniform}(-0.5, 0.5),$ $\ N_1, N_2, N_3 \sim \operatorname{Exp}(0.1)$, where $a$ ranges from $1$ to $10$.
\end{itemize}

\begin{figure*}[tbp]
    \centering
    \includegraphics[width=\textwidth]{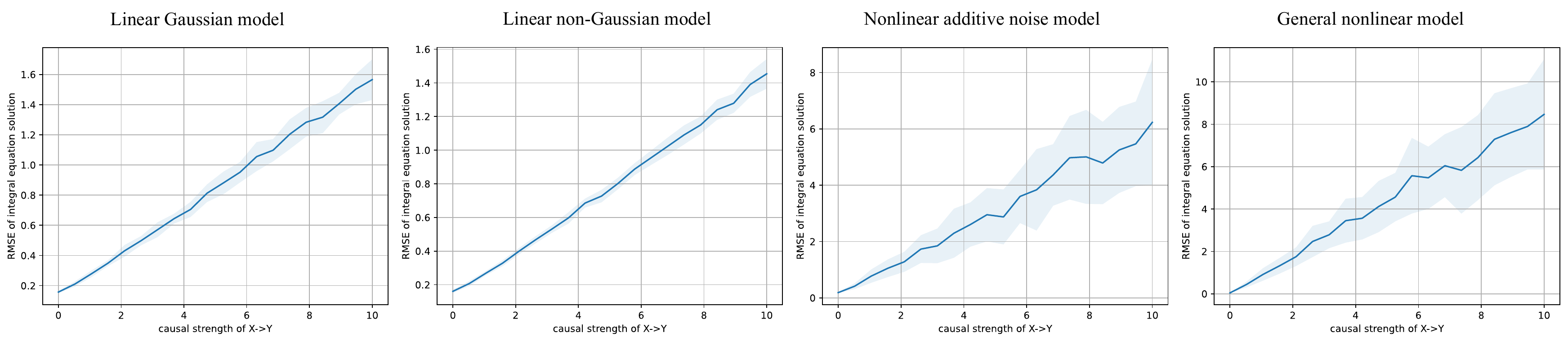}
    \caption{Insolubility of the integral equation when $X\not\ind Y|U$. We consider four examples, namely the linear Gaussian model, the linear nonGaussian model, the nonlinear additive noise model, and the general nonlinear models. For each example, we report the averaged residual of the best fit of the integral equation $f(y|x)=\int h(y,w)f(w|x)dw$ wrt to the correlation strength of $X$ and $Y$. As we can see, when $X$ and $Y$ become strongly correlated ($X\not\ind Y|U$), the residual becomes non-zero, indicating the insolubility of the integral equation. }
    \label{fig.exam-integral-equation-no-solution}
\end{figure*}

For each model, we implement the Maximum Moment Restriction (MMR) with Gaussian kernels in the $\mathrm{mliv}$ package\footnote{\url{https://github.com/causal-machine-learning-lab/mliv}} to find the solution to the conditional moment equation $E[Y-h(W)|X]=0$. The results are shown in Fig.~\ref{fig.exam-integral-equation-no-solution}. As we can see, when $X$ and $Y$ becomes strongly correlated (\emph{i.e.}, $X\not\ind Y|U$ holds), there is no valid solution to the conditional moment function, which indicates the insolubility of the integral equation.

Nonetheless, this condition may not constantly hold given $X\not\ind Y|U$, which means $\mathbb{H}_1$ is indeed a local alternative. Specifically, denote $\left(\lambda_n, \phi_n, \psi_n\right)_{n=1}^{\infty}$ a singular value decomposition of $K$, we then have the following theorem:

\begin{theorem}
    Suppose that the following conditions hold:
    \begin{enumerate}
    \itemsep0em
        \item Completeness of $\mathcal{L}_{X|U}$.
        \item $\iint f(u|x) f(x|u) dx du<\infty$.
        \item $\int f^2(y|x) f(x) dx<\infty$.
        \item $\sum_{n=1}^{\infty} \lambda_n^{-2}\left|\left\langle f(y|x), \psi_n\right\rangle\right|<+\infty$, where $\left\langle f(y|x), \psi_n\right\rangle:=\int f^2(y|x) \psi_n(x) f(x) d x$.
    \end{enumerate}
Then there exists $h(u, y)$ such that $\int h(u, y) f(u|\mathrm{x})=f(y| x)$ holds for all $(x, y)$.
\end{theorem}

\begin{remark}
    These conditions should be compatible with our completeness condition in Asm.~\ref{asm.comp-main}. Particularly, the completeness of $\mathcal{L}(U|X)$ and that of $\mathcal{L}(X|U)$ roughly means the $U$ and $X$ have similar levels of variation. To explain, note that if $U$ and $X$ are discrete, then we have $P(U|X)$ and $P(X|U)$ are both invertible. 
\end{remark}

\begin{remark}
    This result, which means Asm.~\ref{asm.int-eq-main} does not always hold, means it is reasonable to take Asm.~\ref{asm.int-eq-main} as a condition, instead of a theorem or a claim/hypothesis. 
\end{remark}

\begin{proof}
    Simply follow the proof of Prop. 1 of \cite{miao2018identifying} with the following replacements:
    \begin{itemize}
        \item replace $f(w | z, x)$ with $f(u | x)$,
        \item replace $f(z | w, x)$ with $f(x | u)$,
        \item replace $f(y | z, x)$ with $f(y | x)$,
        \item replace $\left(K_x, L^2(f(w | x)), L^2(f(z | x)), \lambda_{x, n}, \phi_{x, n}, \psi_{x, n}\right)$ with $\left(K, L^2(f(u)), L^2(f(x)), \lambda_n, \phi_n, \psi_n\right)$.
    \end{itemize}
\end{proof}

\subsection{Proof of Thm.~\ref{thm.type-ii-main}: Asymptotic power of the test}
\label{app-sec.proof-type-ii}

\textbf{Theorem~\ref{thm.type-ii-main}.} \emph{Suppose that Asm.~\ref{asm.comp-main},~\ref{asm.tv-smooth-main},~\ref{asm.tight-main},~\ref{asm.int-eq-main} hold, then for any $\alpha\in (0,1)$, it holds that:
\begin{equation*}
    \lim_{n, l\to\infty}P(T_{n,l}>z_{1-\alpha})=1.
\end{equation*}
In other words, the limiting power of our test is one.}

\begin{remark}
    According to the following proof, we have that under the same condition, there exists a solution $h$ such that $f(y|x) = \int h(y,w)f(w|x)dw$ under $\mathbb{H}_0$. 
\end{remark}

\begin{proof}
    First recall that we use the notations $\{\mathcal{X}_i\}_{i=1}^I,\{\mathcal{Y}_j\}_{j=1}^J,\{\mathcal{U}_k\}_{k=1}^K,\{\mathcal{W}_l\}_{l=1}^L$ to denote the partitions of $\mathcal{X},\mathcal{Y},\mathcal{U},$ and $\mathcal{W}$, respectively. Moreover, we use $l$ and $L$ interchangeably. We denote $q:=P(\tilde{y}|\tilde{X})^{T}$ and $Q:=P(\tilde{W}|\tilde{X})$. The proof consists of two parts. In \textbf{part I}, we show that under $\mathbb{H}_1$, $q\not \in \mathrm{im}(Q)$ as $L\to\infty$. Then, in \textbf{part II}, given that $q\not \in \mathrm{im}(Q)$, we show $P(T> z_{1-\alpha})$ can be controlled to zero.

    \textbf{Part I.} We show that under $\mathbb{H}_1$, $q\not \in \mathrm{im}(Q)$ as $L\to\infty$. To see this, we first show that there exists $L_0$ such that as long as $I>K=L>L_0$, $q\not \in \mathrm{im}(Q)$. Specifically, let 
    \begin{align*}
        \mathrm{res}(h):=\iint |f(y|x)-\int h(y,w)f(w|x)dw|dy f(x) dx,
    \end{align*}
    and let $h_0:=\arg\min_h \mathrm{res}(h)$. Since $f(y|x)\geq 0$ and $f(w|x)\geq 0$, since $h_0$ minimizes $\mathrm{res}(h)$, we have $h_0\geq 0$ a.s. Besides, we have the following by taking $M$ sufficiently large, 
    \begin{align*}
        \int_{|x|\leq M} \int_{|y|\leq M} \left|f(y|x)-\int h(y,w)f(w|x)dw \right|dy f(x) dx > \frac{1}{2} \epsilon_0, \text{ for any $h\geq 0$}. 
    \end{align*}
     To see this, note that for any $M$, we have $\mathrm{res}(h)=\mathrm{res}_{\leq M}{h} + \mathrm{res}_{>M}(h)$, where 
     \begin{align*}
         \mathrm{res}_{\leq M}(h) & :=\int_{|x|\leq M} \int_{|y|\leq M} |f(y|x)-\int h(y,w)f(w|x)dw|dy f(x) dx \\
         \mathrm{res}_{> M}(h) & :=\left(\int_{|x|> M} \int_{|y|\leq M}+\int_{|x|\leq M} \int_{|y|> M}+\int_{|x|> M} \int_{|y|> M} \right) \left|f(y|x)-\int h(y,w)f(w|x)dw \right|dy f(x) dx.
     \end{align*}
     For $\mathrm{res}_{>M}(h)$, we have 
     \begin{align*}
         \mathrm{res}_{>M}(h) & \geq \left(\int_{|x|> M} \int_{|y|\leq M}+\int_{|x|\leq M} \int_{|y|> M}+\int_{|x|> M} \int_{|y|> M} \right) f(y|x)f(x)dydx \\
         & = P(|X|>M)+P(|X|\leq M,|Y|>M) < \frac{1}{2}\epsilon_0
     \end{align*}
     if $M$ is sufficiently large. Then, without loss of generality, we have $\mathrm{res}(h)>\epsilon_0$ for some $\epsilon_0$ and $M > 0$ such that $|X| \leq M, |Y|\leq M$. Since $f(y|x)$ is continuous, there exists $M_0$ such that $f(y|x)\leq M_0$ on $\{(x,y):|x|\leq M, |y|\leq M\}$. Therefore, we have $0\leq h_0\leq M_0$, which means we only need to consider $h$ with $0\leq h\leq M_0$.

    Further, there exists $M_w>0$ such that 
    \begin{align*}
        \int_{|x|\leq M}\int_{|y|\leq M} \left|f(y|x)-\int_{|w|\leq M_w} h(w,y) f(w|x) dw \right| dyf(x)dx >\frac{3}{4}\epsilon_0
    \end{align*}
    for any $0\leq h \leq M_0$. To see this, note that 
    \begin{align*}
        \mathrm{res}(h) & =\int_{|x|\leq M} \int_{|y|\leq M} \left|f(y|x)-\int_{|w|\leq M_w} h(y,w)f(w|x)dw\right|dy f(x) dx \\
        & \quad \quad \quad \quad \quad \quad \quad \quad + \int_{|x|\leq M}\int_{|y|\leq M} \int_{|w|>M_w} h(w,y)f(w|x)dwdyf(x)dx \\
        & \leq \int_{|x|\leq M} \int_{|y|\leq M} |f(y|x)-\int_{|w|\leq M_w} h(y,w)f(w|x)dw|dy f(x) dx + M\cdot M_0 P(|W|>M_w,|X|\leq M) \\
        & \leq \int_{|x|\leq M} \int_{|y|\leq M} |f(y|x)-\int_{|w|\leq M_w} h(y,w)f(w|x)dw|dy f(x) dx + M\cdot M_0 P(|W|>M_w).
    \end{align*}
    Therefore, for any $0\leq h\leq M_0$, we have $\mathrm{res}^{M_w}(h)>\frac{3}{4}\epsilon_0$. Note that 
    \begin{align*}
        \mathrm{res}^{M_w}(h)=\sum_{i=1}^I \sum_{j=1}^J \int_{\mathcal{X}_i} \int_{\mathcal{Y}_j} \left|f(y|x)-\int_{|w|\leq M_w} h(w,y) f(w|x) dw \right|dyf(x)dx.
    \end{align*}
    Since $f(y|x)$ and $f(w|x)$ are respectively uniformly continuous in $\{(x,y):|x|\leq M, |y|\leq M\}$ and $\{(x,w):|x|\leq M, |w|\leq M_w\}$, there exists $I_{0,1}$ $J_{0,1}$ such that for any $I>I_{0,1}, J>J_{0,1}$, we have 
    \begin{align}
    \label{eq.bound-01}
        \left|\mathrm{res}^{M_w}(h)-\sum_{i=1}^I \sum_{j=1}^J\left|f(\bar{y}_j|\bar{x}_i)-\int_{|w|\leq M_w} h(w,y) f(w|\bar{x}_i)dw\right| P(X \in \mathcal{X}_i)\mathrm{len}(\mathcal{Y}_j) \right| < \frac{1}{4} \epsilon_0
    \end{align}
    for any $\bar{x}_i \in \mathcal{X}_i, \bar{y}_j \in \mathcal{Y}_j$, and $h$ that is constant for any $w$ and $y\in\mathcal{Y}_j$. To see this, note that $f(y|x)$ and $f(w|x)$ are uniformly continuous. Then $\forall\epsilon>0$, there exists $\delta>0$ such that 
    \begin{align*}
        \left|f(w_1|x_1)-f(w_2|x_2) \right| \leq \epsilon, \
        \left|f(y_1|x_1)-f(y_2|x_2)\right|\leq \epsilon
    \end{align*}
    for any $\sqrt{(w_1-w_2)^2+(x_1+x_2)^2}<\delta$ and $\sqrt{(y_1-y_2)^2+(x_1+x_2)^2}<\delta$. Therefore, as long as the partition satisfies that for each $i,j, \mathrm{len}(\mathcal{X}_i)\leq \delta,\mathrm{len}(\mathcal{Y}_j)<\delta$, then for any $h$ that is constant for any $w$ and $y\in\mathcal{Y}_j$, we have
    \begin{align*}
        & \iint_{\mathcal{X}_i,\mathcal{Y}_j}|P(\bar{y}_j|\bar{x}_i)-f(y|x)|f(x) dxdy  \leq \epsilon \delta, \\
        & \iint_{\mathcal{X}_i,\mathcal{Y}_j} \left| \int_{|w|\leq M_w} h(w,\bar{y}_j) f(w|\bar{x}_i) dw - \int_{|w|\leq M_w} h(w,y)f(w|x)dw \right|f(x)dxdy \\
        & \leq \iint_{\mathcal{X}_i,\mathcal{Y}_j}  \int_{|w|\leq M_w} h(w,\bar{y}_j) \left|f(w|\bar{x}_i)-f(w|x)\right| dw f(x)dxdy \leq M_0M_w \epsilon\delta,
    \end{align*}
    for some $\bar{x}_i \in \mathcal{X}_i$ and $\bar{y}_j \in \mathcal{Y}_i$. By taking the $\epsilon$ small enough, we have the desired conclusion.

    Similarly, there exists $I_{0,2},J_{0,2}$ such that as long as $I>I_{0,2},J>J_{0,2}$, we have 
    \begin{align}
        & \left|\sum_{i=1}^I \sum_{j=1}^J  \left| \int_{\mathcal{X}_i}\int_{\mathcal{Y}_j} f(y|x)-\int_{|w|\leq M_w} h(w,y) f(w|x)dw  dyf(x) dx \right|  - \right. \nonumber \\
        & \left. \sum_{i=1}^I \sum_{j=1}^J \left| f(\bar{y}_j|\bar{x}_i) -\int_{|w|\leq M_w} h(w,y)f(w|\bar{x}_i) dw \right| P(X \in \mathcal{X}_i)\mathrm{len}(\mathcal{Y}_j) \right|<\frac{1}{4}\epsilon_0 \label{eq.bound-02}
    \end{align}
   for any $h$ that is constant for any $w$ and $y\in\mathcal{Y}_j$. In addition, by noting that 
   \begin{align*}
       \int_{\mathcal{X}_i}\int_{\mathcal{Y}_j}\left(f(y|x)-\int h(w,y)f(w|x)dw\right)dyf(x)dx=P(\tilde{y}_j|\tilde{x}_i)-\int g_l(w)f(w|X\in\mathcal{X}_i)dw P(X \in \mathcal{X}_i),
   \end{align*}
   where $g_l(w):=\int_{Y\in\mathcal{Y}_j} h(w,y)dy$, we have: 
   \begin{align}
       \sum_{i=1}^I \sum_{j=1}^J \left| \int_{\mathcal{X}_i} \int_{\mathcal{Y}_j} \left(f(y|x)-\int h(w,y)f(w|x)dw\right)dyf(x)dx \right| \nonumber\\
       =\sum_{i=1}^I \sum_{j=1}^J \left| P(\tilde{y}_j|\tilde{x}_i)-\int g_l(w)f(w|X\in\mathcal{X}_i)dw \right| P(X\in \mathcal{X}_i).\label{eq.bound-03}
   \end{align}

    By combining Eq.~\eqref{eq.bound-01},~\eqref{eq.bound-02},~\eqref{eq.bound-03}, for any $I>\max(I_{0,1},I_{0,2})$, $J> \max(J_{0,1},J_{0,2})$, we have: 
    \begin{align*}
        \left|\mathrm{res}^{M_w}(h)-\sum_{i=1}^I \sum_{j=1}^J \left|P(\tilde{y}_j|\tilde{x}_i)-\int g_l(w)f(w|X\in\mathcal{X}_i)dw \right|P(X\in \mathcal{X}_i) \right| \leq \frac{1}{2}\epsilon_0.
    \end{align*}
    Then for any $I,J$, there exist $L_0$ such that for any $L>L_0$, we have: 
    \begin{align*}
        \left|\mathrm{res}^{M_w}(h)-\sum_{i=1}^I \sum_{j=1}^J \left|P(\tilde{y}_j|\tilde{x}_i)-\sum_k P(\tilde{w}_k|\tilde{x}_i) g_l(w)f(w|X\in\mathcal{X}_i)dw \right|P(X\in \mathcal{X}_i) \right| \leq \frac{3}{4}\epsilon_0.
    \end{align*}
    If $q=Q^T c$ for some $c=[[c_{1,1},...,c_{L,1}],...,[c_{1,J},...,C_{L,J}]]^T$, with $c_{l,i,j}=c_{l,i^',j}$, then we let $h(w,y)=c_{k,l}$ for $w\in\mathcal{W}_l,y\in\mathcal{Y}_j$ and have $p(\tilde{y}_j|\tilde{x}_i)=\sum_l p(\tilde{w}_k|\tilde{x}_i) c_{l,j}$. Therefore, we have $\sum_{i=1}^I\sum_{j=1}^J |p(\tilde{y}_l|\tilde{x}_i)-\sum_l p(\tilde{w}_l|\tilde{x}_i) g_l(\tilde{u}_l)|P(X\in \mathcal{X}_i)=0$ for some $h$ that is constant in each $\mathcal{Y}_j$. Therefore, we have $\mathrm{res}^{M_w}(h)<\frac{3}{4}\epsilon_0$, which obtains the contradiction.

    \textbf{Part II.} Given that $q\not \in \mathrm{im}(Q)$, we show $\lim_{n,L\to \infty} P_{\mathbb{H}_1}(T\leq z_{1-\alpha})=0$. To show this, we first discuss the distribution of $T$ given $q\not \in \mathrm{im}(Q)$. First, we have $||\Omega \Sigma_y^{-1/2} q_y||^2_2>0$. To see this, we first have that $\mathrm{ker}(\Omega)=\mathrm{im}(\Sigma^{-1/2}Q^T)$. 

    To see this, we have $\forall \vec{a}\in \mathrm{im}(\Sigma^{-1/2}Q^T), \exists \vec{b}, s.t. \vec{a} = \Sigma^{-1/2}Q^T \vec{b}$. According to the definition of $\Omega$ in Eq.~\eqref{eq.omega}, we have $\Omega \vec{a} = \Omega \Sigma^{-1/2}Q^T \vec{b} = \vec{0}$, which means $\mathrm{im}(\Sigma^{-1/2}Q^T) \subseteq \mathrm{ker}(\Omega)$. What is left is to show $\mathrm{dim}(\mathrm{im}(\Sigma^{-1/2}Q^T)) = \mathrm{dim}(\mathrm{ker}(\Omega))$. First, we have $\mathrm{dim}(\mathrm{im}(\Sigma^{-1/2}Q^T)) = \mathrm{rank}(\Sigma^{-1/2}Q^T) = L(J-1)$. Since $\mathrm{rank}(\Omega)=(I-L)(J-1)$ and $\Omega \in \mathbb{R}^{I(J-1) \times I(J-1)}$, we have $\mathrm{nul}(\Omega)=I*(J-1)-(I-L)(J-1)=L(J-1)$ ($\mathrm{nul}(\Omega)$ is the null space of $\Omega$) and therefore $\mathrm{dim}(\mathrm{ker}(\Omega))=L(J-1)=\mathrm{dim}(\mathrm{im}(\Sigma^{-1/2}Q^T))$. Therefore, we have $\mathrm{ker}(\Omega)=\mathrm{im}(\Sigma^{-1/2}Q^T)$.

    Therefore, $q\not \in \mathrm{im}(Q)$ means $\Sigma^{-1/2} q \not \in \mathrm{img}(\Sigma^{-1/2} Q^T) = \mathrm{ker}(\Omega)$, and we have $\Omega\Sigma^{-1/2} q \neq \vec{0}$ and $|| \Omega\Sigma^{-1/2} q ||^2_2>0$. Second, let $\vec{c} := \Omega \Sigma^{-1/2} q$, we show $T$ follows the noncentral chi square distribution $\chi^2(r,\lambda)$ with $\lambda=n||\vec{c}||^2_2$ and $r=(I-L)(J-1)$. To be specific, we have $\sqrt{n} (\xi - \vec{c}) \to N(0,\Omega)$ and therefore $\sqrt{n} \xi \to N(\sqrt{n}\vec{c}, \Omega)$. Repeating the orthogonal decomposition trick used in the proof of Theorem~\ref{thm.control.type-i}, we have:
    \begin{align*}
            T := n \xi^T \xi &\to N(\sqrt{n}\vec{c},\Omega)^T N(\sqrt{n}\vec{c},\Omega) = \{V N(\sqrt{n}\vec{c},\Omega)\}^T \{V N(\sqrt{n}\vec{c},\Omega)\} \\
            &= N(\sqrt{n}V\vec{c}, \mathrm{diag}(1,...,1,0,...,0))^T N(\sqrt{n}V\vec{c}, \mathrm{diag}(1,...,1,0,...,0)) = \chi^2(r,\lambda),
    \end{align*}

    Given that $T$ follows the noncentral $\chi^2(r,\lambda)$ distribution as $L\to\infty$, we then show $\lim_{n,L\to\infty}P_{\mathbb{H}_1}(T\leq z_{1-\alpha})=0$, which is equivalent to show that $\forall \epsilon>0, \exists N, \forall n\geq N, P_{\mathbb{H}_1}(T_y\leq z_{1-\alpha}) \leq\epsilon$. In details, denote $F(x;r,\lambda)$ the cdf of $\chi^2(r,\lambda)$ and let $f(\lambda):=F(x;r,\lambda)$. We first show for fixed ($x,r$) (the $x$ is $z_{1-\alpha}$ in our problem), $\lim_{\lambda\to\infty} f(\lambda) = 0$. According to \cite{nuttall1972some}, we can represent the cdf with Marcum Q function, \emph{i.e.}, $F(\lambda;x,r) = 1 - Q_{r/2}(\sqrt{\lambda},\sqrt{x})$, where
    \begin{align*}
        &Q_v(a,b) = 1 + H_v(a,b) + C_v(a,b) \\
        &H_v(a,b) = \frac{(a/b)^{1-v}}{2\pi} \exp\left(-\frac{1}{2}(a^2+b^2)\right) \int_{0}^{2\pi} \frac{\mathrm{cos}(v-1)\theta - (a/b)\mathrm{cos}(v\theta)}{1-2(a/b)\mathrm{cos}\theta + (a/b)^2} \exp(ab\mathrm{cos}\theta)\mathrm{d}\theta, \\
        &C_v(a,b) = \frac{\mathrm{sin}(v\pi)}{\pi} \exp\left(-\frac{1}{2}(a^2+b^2)\right) \int_0^1 \frac{x^{v-1} (b/a)^{v-1}}{(a/b)+x} \exp(-\frac{ab}{2}(x+\frac{1}{x})) \mathrm{d}x.
    \end{align*}

    Therefore, to show $\lim_{\lambda\to\infty} f(\lambda) = 0$, we show $\lim_{\lambda\to\infty} Q_{k/2}(\sqrt{\lambda},\sqrt{x}) = 0$, which needs $\lim_{a\to\infty} H_v(a,b) = 0$ and $\lim_{a\to\infty} C_v(a,b) = 0$. To see $\lim_{a\to\infty} H_v(a,b) = 0$, note that:
    \begin{align*}
        &\lim_{a\to \infty} \frac{(a/b)^{1-v}}{2\pi} = 0, \quad \hfill\#\, v \,\text{is a large number, so}\, 1-v<0, \\
        &\lim_{a\to \infty} \exp\left(-\frac{1}{2}(a^2+b^2)\right) \exp (ab\mathrm{cos}\theta) = 0, \quad \hfill\#\,\text{because}\, \lim_{a\to \infty}a^2-2ab\mathrm{cos}\theta+b^2 = \infty, \\
        &\lim_{a\to \infty} \frac{\mathrm{cos}(v-1)\theta-(a/b)\mathrm{cos}(v\theta)}{1-2(a/b)\mathrm{cos}\theta+(a/b)^2} = 0 \quad \#\, \text{this is like}\, \lim_{a\to \infty} \frac{a}{1-a+a^2}= 0.
    \end{align*}

    Therefore, use Lebesgue’s dominated convergence theorem (our integrand can be dominated by $1$ when $a$ is sufficiently large, and $1$ is an integrable function) to exchange the limitation and integration, we have $\lim_{a\to\infty} H_v(a,b) = 0$. A similar process can show $\lim_{a\to\infty} C_v(a,b) = 0$. Now, with $\lim_{\lambda\to\infty} f(\lambda) = 0$, we have $\forall \epsilon>0,\exists\, \Lambda,\forall\lambda>\Lambda, f(\lambda):=P_{\chi^2(r,\lambda)}(T_y\leq z_{1-\alpha})\leq \epsilon$. Therefore, we have $\forall \epsilon>0,\exists N:=\frac{\Lambda}{||\vec{c}||^2_2},\forall n\geq N,\lambda:=n||\vec{c}||^2_2 \geq \Lambda$ and hence $P_{\mathbb{H}_1}(T_y\leq z_{1-\alpha})\leq \epsilon$, which concludes the proof.
\end{proof}


\end{document}